\documentclass{sig-alternate}
\usepackage[english]{babel}
\usepackage[latin1]{inputenc}
\usepackage{amssymb}
\usepackage[ruled]{algorithm2e}
\usepackage{algpseudocode}
\usepackage{bm}
\usepackage{amsmath}
\usepackage{graphicx}
\usepackage{epsfig}
\usepackage{url}
\usepackage{psfrag}
\usepackage{balance}
\usepackage{pstricks}
\usepackage{times}
\usepackage{array}
\usepackage{subfigure}
\usepackage{float}
\usepackage{multirow}

\newcommand{\transpose}{^{\ensuremath{\mathsf{T}}}}

\begin{document}

\title{Minfer: Inferring Motif Statistics From Sampled Edges}
\numberofauthors{1}
\author{
\alignauthor
Pinghui Wang$^{1}$, John C.S.\ Lui$^{2}$, and Don Towsley$^{3}$\\
{\small
\affaddr{$^{1}$Noah's Ark Lab, Huawei, Hong Kong}\\
\affaddr{$^{2}$Department of Computer Science and Engineering, The Chinese University of Hong Kong, Hong Kong}\\
\affaddr{$^{3}$Department of Computer Science, University of Massachusetts Amherst, MA, USA}\\
\{wang.pinghui\}@huawei.com, cslui@cse.cuhk.edu.hk, towsley@cs.umass.edu
}
}

\maketitle

\begin{abstract}
Characterizing motif (i.e., locally connected subgraph patterns) statistics is important
for understanding complex networks such as online social networks
and communication networks.
%Previous work assumes the graph topology of interest is known,
Previous work made the strong assumption that the graph topology of interest
is known,
and that the dataset either fits into main memory or is stored on disk
%and can be fitted into memory.
%Or, the graph is well organized and stored on disks (e.g., stored in relational databases)
such that it is not expensive to obtain all neighbors of any given node.
%In practice, due to limited resources (e.g., it is very expensive to collect and store all connections on backbone routers), however, the graph of interest might be unknown and the available data is a graph sampled from it.
In practice, researchers have to deal with the situation where the graph topology is
unknown, either because the graph is dynamic, or because
it is expensive to collect and store all topological and meta information on disk.
Hence, what is available to researchers is only a snapshot of the graph generated by sampling edges from the graph at random,
which we called a ``{\em RESampled graph}''.
%Or, when the large graph is given in a wild format and it is expensive to preprocess and organize it on disk,
%one might want to quickly get an approximate answer by computing the motif statistics of a relatively small sampled graph in memory.
Clearly, a RESampled graph's motif statistics may be quite different from the
underlying original graph.
To solve this challenge, we propose a
%method
framework and implement a system called Minfer,
which can take the given RESampled graph and accurately
infer the underlying graph's motif statistics.
We also use Fisher information to bound the errors of our estimates.
Experiments using
%widely known datasets,
large scale datasets show our method to be accurate.

\end{abstract}

%!TEX root = samplingmotifs.tex
\section{Introduction} \label{sec:introduction}
Complex networks are widely studied across many fields of science and technology,
from physics to biology, and from nature to society.
Networks which have similar global topological features
such as degree distribution and graph diameter
can exhibit significant differences in their local structures.
%Therefore, there is considerable interest in exploring small
There is a growing interest to explore these local structures
(also known as ``{\em motifs}''),
which are small connected subgraph patterns that
form during the growth of a network.
%connected subgraph patterns in networks,
%which are often shaped during their growth
%Local structures
Motifs have many applications, for example, they are used to
%and have been used to
characterize communication and evolution patterns
in online social networks (OSNs)~\cite{ChunIMC2008,Kunegis2009,ZhaoNetsci2011,
Ugander2013},
pattern recognition in gene expression
profiling~\cite{Shenorr2002}, protein-protein interaction
prediction~\cite{Albert2004}, and coarse-grained topology
generation of networks~\cite{Itzkovitz2005}.
For instance, 3-node motifs such as
``{\em the friend of my friend is my friend}''
and ``{\em the enemy of my enemy is my friend}''
are well known evolution patterns in signed (i.e., friend/foe) social networks.
Kunegis et al.~\cite{Kunegis2009} considered the significance of
motifs in Slashdot Zoo\footnote{www.slashdot.org} and how they
impact the stability of signed networks.
Other more complex examples include 4-node motifs such as bi-fans and bi-parallels defined in~\cite{Milo2002}.

Although motifs are important characteristics to help researchers to understand
the underlying network,
one major technical hurdle is that it is
computationally expensive to compute motif frequencies
since this requires one to enumerate and count all subgraphs in a network,
and there exist a large number of subgraphs even for a medium size network
with less than one million edges.
For example, the graphs Slashdot~\cite{LeskovecIM2009} and Epinions~\cite{Richardson2003}, which contain approximately $1.0\times 10^5$ nodes and $1.0\times 10^6$ edges have more than $2.0\times 10^{10}$ 4-node connected and induced subgraphs~\cite{TKDDWang2014}.
To address this problem, several sampling methods
have been proposed to estimate the frequency distribution of motifs~\cite{Kashtan2004,Wernicke2006,Bhuiyan2012,TKDDWang2014}.
All these methods require that the entire graph topology fit into memory,
or the existence of an I/O efficient neighbor query API available so
that one can explore the graph topology, which is stored on disk.
%or that a neighbor query friendly API be available to query the graph stored in databases on disk.
In summary,previous work focuses on designing \textbf{\emph{computationally efficient}} methods to characterize motifs when the \textbf{\emph{entire graph}} of interest is given.

\begin{figure}[htb]
\begin{center}
\includegraphics[width=0.4\textwidth]{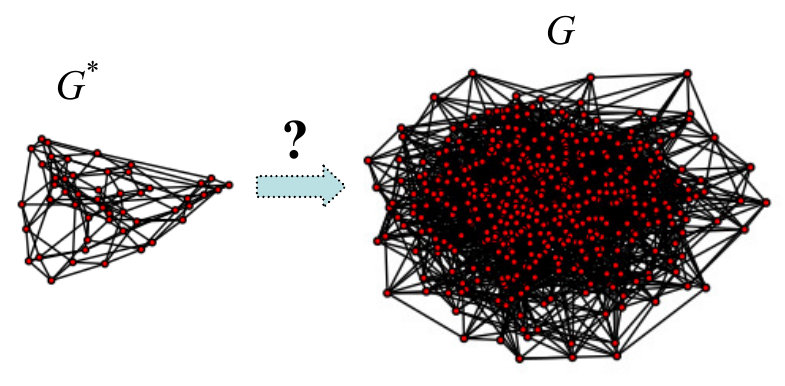}
\caption{An example of the available RESampled $G^*$ and the underlying graph $G$.}
\label{fig:exampleGstar}
\end{center}
\end{figure}

\begin{figure}[htb]
\begin{center}
\includegraphics[width=0.38\textwidth]{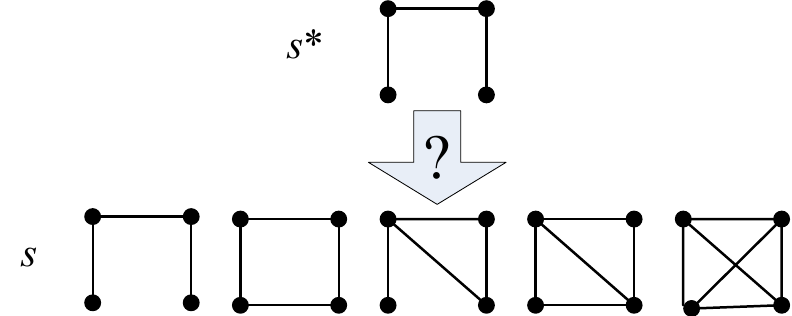}
\caption{$s^*$ is a 4-node induced subgraph in the RESampled graph $G^*$, and $s$ is the original induced subgraph of $s^*$ in the original graph $G$.}
\label{fig:biasexample}
\end{center}
\end{figure}

In practice the graph of interest may not be known,
%either because the graph is dynamic, or because it is expensive to
%collect and query all topological and meta information on disks.
but instead the available dataset is a subgraph sampled from the original graph.
This can be due to the following reasons:
\begin{itemize}
               \item \textbf{Data collection}: Sampling is inevitable for collecting a large dynamic graph given as a high speed stream of edges.
               For example,
               sampling is used to collect network traffic on backbone routers
               in order to study the network graph where a node in the graph represents a host and an edge $(u,v)$ represents a connection from host $u$ to host $v$,
               because capturing the entire traffic is prohibited due to the high speed traffic and limited resources (e.g. memory and computation) of network devices.
               \item \textbf{Data transportation}: Sampling may also be required to reduce the high cost of transporting an entire dataset to a remote data analysis center.
               \item \textbf{Memory and computation}: Sometimes the graph of interest is given in a memory expensive format such as a raw text file, and may be too large to fit into memory.
                   Moreover, it may be too expensive to preprocess and organize it on disk.
                    In such cases,
                    it may be useful to build a relatively small graph consisting of edges sampled
                    from the graph file at random,
                    and compute its motif statistics in memory.
\end{itemize}
A simple example is given in Fig.~\ref{fig:exampleGstar},
where the sampled graph $G^*$ is derived from the dataset representing $G$.
In this work,
%For simplicity,
we assume the available graph $G^*$ is obtained through random edge sampling (i.e, each edge is independently sampled with the same probability $0\le p\le 1$),
which is popular and easy to implement in practice.
Formally, we denote the graph $G^*$ as a RESampled graph of $G$.
One can easily see that a RESampled graph's motif statistics will differ from
those of the original graph due to uncertainties introduced by sampling.
For example, Fig.~\ref{fig:biasexample} shows that
$s^*$  is a 4-node induced subgraph in the RESampled graph $G^*$, and we do not know
from which original induced subgraph $s$ in $G$ that it derives.
$s$ could be any one of the five subgraphs depicted in Fig.~\ref{fig:biasexample}.

\begin{figure}[htb]
\begin{center}
\includegraphics[width=0.5\textwidth]{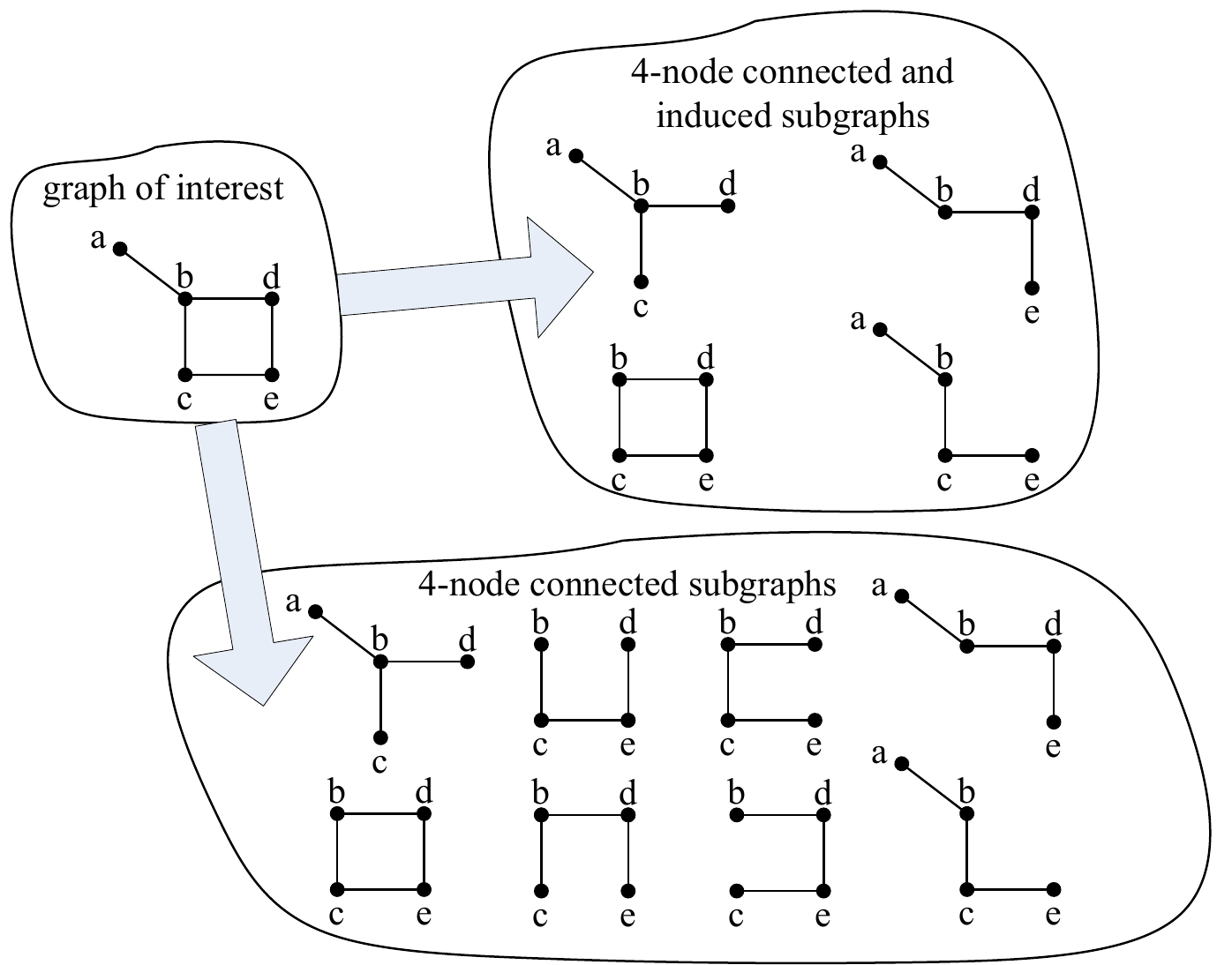}
\caption{4-node CISes vs. 4-node connected subgraphs.}
\label{fig:inducdvsnoninduced}
\end{center}
\end{figure}

Unlike previous
methods~\cite{Kashtan2004,Wernicke2006,Bhuiyan2012,TKDDWang2014},
we aim to design an \textbf{\emph{accurate method}} to infer motif
statistics of the original graph $G$ from the available RESampled graph $G^*$.
These previous methods focus on designing computationally efficient sampling methods
based on sampling \emph{induced} subgraphs in $G$ to avoid the problem shown in Fig.~\ref{fig:biasexample}.
Hence they fail to infer the underlying graph's motif statistics from the given RESampled graph.
The gSH method in ~\cite{AhmedKDD2014} can be used to estimate the number of connected subgraphs from sampled edges.
However it cannot be applied to characterize motifs, i.e., connected and \textbf{\emph{induced}} subgraphs (or CISes),
because motif statistics can differ from connected subgraphs' statistics.
For example,
Fig.~\ref{fig:inducdvsnoninduced} shows that $75\%$ of a graph's 4-node connected subgraphs  are isomorphic to a 4-node line (i.e., the {\em first} motif in Fig.~\ref{fig:345nodeclasses} (b)),
while $50\%$ of its 4-node CISes are isomorphic to a 4-node line.

{\bf Contribution:} Our contribution can be summarized as:
To the best of our knowledge, we are the first to study and provide
an accurate and efficient solution to estimate motif statistics from a given RESampled graph.
%to the above problem.
We introduce a probabilistic model to study
the relationship between motifs in the RESampled graph and in the underlying graph.
Based on this model, we propose an accurate method, Minfer,
to infer the underlying graph's motif statistics from the RESampled graph.
We also provide a Fisher information based method to bound the error of our estimates.
Experiments on a variety of real world datasets show that our method can
accurately estimate the motif statistics of a graph
based on a small RESampled graph.

This paper is organized as follows: The problem formulation is presented in Section~\ref{sec:problem}. Section~\ref{sec:methods} presents our method  (i.e. Minfer) for inferring subgraph class concentrations of the graph under study from a given RESampled graph.
Section~\ref{sec:enumerations} presents methods for computing the given RESampled graph's motif statistics.
The performance evaluation and testing results are presented in Section~\ref{sec:results}.
Section~\ref{sec:related} summarizes related work. Concluding remarks then follow.

\begin{table}[htb]
\begin{center}
\caption{Table of notations.\label{tab:notations}}
\begin{tabular}{|c|l|} \hline
%\multirow{2}{*}{$r_t=(u_t, v_t, l_t)$}& stream of records, where $l_t$ is a label\\
%&associated with the edge $(u_t,v_t)$\\ \hline
$G$&$G=(V, E, L)$ is the graph under study\\ \hline
$G^*$&$G^*=(V^*, E^*, L^*)$ is a RESampled graph\\ \hline
%$d(v), v\in V$& degree of a node $v$ in $G$\\ \hline
%$N(v), v\in V$& neighbors of a node $v$ in $G$\\ \hline
%$d^*(v), v\in V^*$& degree of a node $v$ in $G^*$\\ \hline
%$N^*(v), v\in V^*$& neighbors of a node $v$ in $G^*$\\ \hline
$V(s), s\in C^{(k)}$&set of nodes for $k$-node CIS $s$ \\ \hline
$E(s), s\in C^{(k)}$&set of edges for $k$-node CIS $s$\\ \hline
$M(s)$&associated motif of CIS $s$\\ \hline
$T_k$&number of $k$-node motif classes\\ \hline
$M_i^{(k)}$&$i$-th $k$-node motif\\ \hline
$C^{(k)}$&set of $k$-node CISes in $G$\\ \hline
$C_i^{(k)}$&set of CISes in $G$ isomorphic to $M_i^{(k)}$\\ \hline
$n^{(k)}=|C^{(k)}|$&number of $k$-node CISes in $G$\\ \hline
$n_i^{(k)}=|C_i^{(k)}|$&number of CISes in $G$ isomorphic to $M_i^{(k)}$\\ \hline
$m_i^{(k)}$ & number of CISes in $G^*$  isomorphic to $M_i^{(k)}$\\ \hline
$\omega_i^{(k)}=\frac{n_i^{(k)}}{n^{(k)}}$&concentration of motif $M_i^{(k)}$ in $G$\\ \hline
$P$ & matrix $P=[P_{ij}]_{1\le i,j\le T_k}$\\ \hline
\multirow{3}{*}{$P_{i,j}$}&probability that a $k$-node CIS $s^*$ in $G^*$ \\
& isomorphic to $M_i^{(k)}$ given its original\\
& CIS $s$ in $G$ isomorphic to $M_j^{(k)}$\\ \hline
\multirow{2}{*}{$\phi_{i,j}$} &number of subgraphs of $M_j^{(k)}$ isomorphic\\
&to $M_i^{(k)}$\\ \hline
$\mathbf{n}^{(k)}$&$\mathbf{n}^{(k)}=(n_1^{(k)}, \ldots, n_{T_k}^{(k)})\transpose$\\ \hline
$\mathbf{m}^{(k)}$&$\mathbf{m}^{(k)}=(m_1^{(k)}, \ldots, m_{T_k}^{(k)})\transpose$\\ \hline
$m^{(k)}$&$m^{(k)}=\sum_{i=1}^{T_k} m_i^{(k)}$\\ \hline
$\rho_i^{(k)}=\frac{m_i^{(k)}}{m^{(k)}}$&concentration of motif $M_i^{(k)}$ in $G^*$\\ \hline
$p$ & probability of sampling an edge\\ \hline
$q$ & $q=1-p$\\ \hline
\end{tabular}
\end{center}
\end{table}

%!TEX root = samplingmotifs.tex
\section{Problem Formulation} \label{sec:problem}
In this section, we first introduce the concept of motif concentration,
then we discuss the challenges of computing motif concentrations in practice.

Denote the underlying graph of interest as a labeled
undirected graph $G=(V, E, L)$, where $V$ is a set of nodes,
$E$ is a set of \emph{undirected} edges, $E\in V\times V$,
and $L$ is a set of labels $l_{u,v}$ associated with edges $(u,v)\in E$.
For example, we attach a label
$l_{u,v}\in \{\to, \leftarrow, \leftrightarrow\}$ to
%each edge that
indicate the direction of the edge $(u,v)\in E$ for a directed network.
Edges may have other labels too, for instance,
in a signed network, edges have positive or negative labels
to represent {\em friend} or {\em foe} relationship.
If $L$ is empty, then $G$ is an unlabeled undirected graph,
which is equivalent to the regular undirected graph.

\begin{figure*}[htb]
\center
\subfigure[3-node]{
\includegraphics[width=0.17\textwidth]{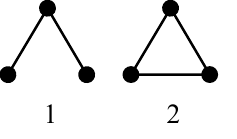}}
\subfigure[4-node]{
\includegraphics[width=0.52\textwidth]{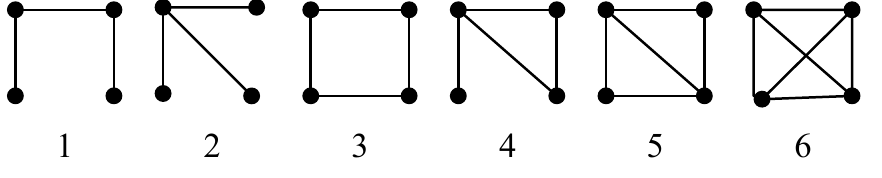}}
\subfigure[5-node]{
\includegraphics[width=\textwidth]{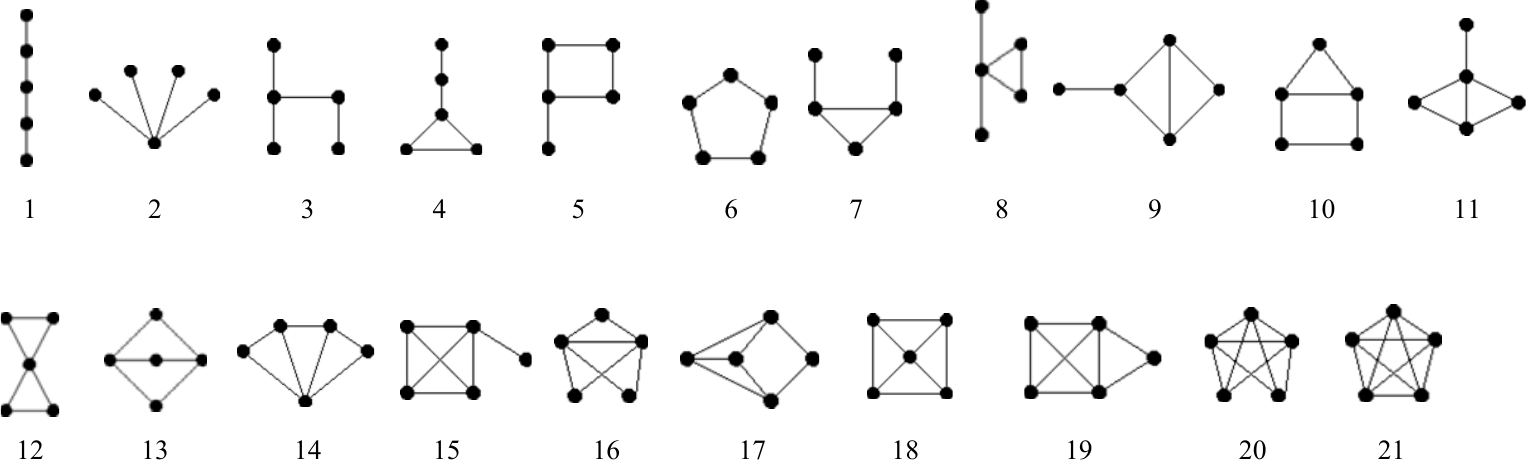}}
\caption{All classes of three-node, four-node, and five-node undirected and connected motifs
         (The numbers are the motif IDs).}
\label{fig:345nodeclasses}
\end{figure*}

\begin{figure*}[htb]
\begin{center}
\includegraphics[width=0.99\textwidth]{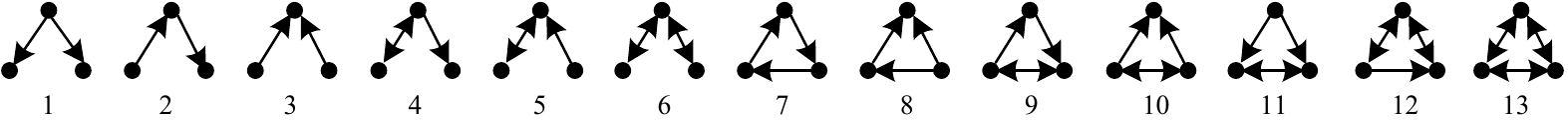}
\caption{All classes of three-node directed and connected motifs (The numbers are the motif IDs).}
\label{fig:3nodeclasses}
\end{center}
\end{figure*}

\begin{figure}[htb]
\begin{center}
\includegraphics[width=0.45\textwidth]{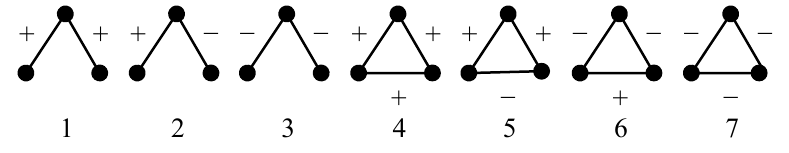}
\caption{All classes of three-node signed and undirected motifs (The numbers are the motif IDs).}
\label{fig:signclasses3nodes}
\end{center}
\end{figure}

Motif concentration is a metric that
represents the distribution of various subgraph patterns that appear in $G$.
To illustrate, we show the 3-, 4- and 5-nodes subgraph patterns in
Figs.~\ref{fig:345nodeclasses},~\ref{fig:3nodeclasses},and~\ref{fig:signclasses3nodes}
respectively.
To define motif concentration formally, we first need to introduce some notation.
For ease of presentation,
Table~\ref{tab:notations} depicts the notation used in this paper.
%used throughout the paper in Table~\ref{tab:notations}.

An induced subgraph of $G$, $G'=(V', E', L')$, $V'\subset V$, $E'\subset E$ and $L'\subset L$,
is a subgraph whose edges and associated labels are all in $G$, i.e. $E'=\{(u, v): u, v \in V', (u,v)\in E\}$,
$L'=\{l_{u,v}: u, v \in V', (u, v)\in E\}$.
We define $C^{(k)}$ as the set of all connected induced subgraphs (CISes) with $k$ nodes in $G$,
and denote $n^{(k)}=|C^{(k)}|$.
For example, Fig.~\ref{fig:inducdvsnoninduced} depicts all possible 4-node CISes.
Let $T_k$ denote the number of $k$-node motifs and
$M_i^{(k)}$ denote the $i^{th}$ $k$-node motif.
For example, $T_4 = 6$ and $M_1^{(4)}, \ldots, M_6^{(4)}$ are the six
4-node undirected motifs depicted in Fig.~\ref{fig:345nodeclasses} (b).
Then we partition $C^{(k)}$ into $T_k$ equivalence classes,
or $C_1^{(k)}, \ldots, C_{T_k}^{(k)}$,
where CISes within $C_i^{(k)}$ are isomorphic to $M_i^{(k)}$.

Let $n_i^{(k)}$ denote the frequency of the motif $M_i^{(k)}$, i.e., the number of the CISes in $G$ isomorphic to $M_i^{(k)}$.
Formally, we have $n_i^{(k)}=|C_i^{(k)}|$, which is the number of CISes in $C_i^{(k)}$.
Then the concentration of $M_i^{(k)}$ is defined as
\begin{equation*}%\label{eq:concentration}
\omega_i^{(k)}=\frac{n_i^{(k)}}{n^{(k)}}, \qquad 1\le i\le T_k.
\end{equation*}
Thus, $\omega_i^{(k)}$ is the {\em fraction} of $k$-node CISes isomorphic to the motif $M_i^{(k)}$ among all $k$-node CISes.
In this paper, we make the follow assumptions:
\begin{itemize}
\item \textbf{Assumption 1}: The complete $G$ is not available
to us, but a RESampled graph $G^*=(V^*, E^*, L^*)$ of $G$ is given,
      where $V^*\in V$, $E^*\in E$, and $L^*$ are node, edge, and edge label sets of $G^*$ respectively.
$G^*$ is obtained by random edge sampling, i.e., each edge in $E$ is independently sampled
with the same probability $0< p <1$,
      where $p$ is known in advance.
\item \textbf{Assumption 2}: The label of a sampled edge $(u,v)\in G^*$ is the same as that of $(u,v)$ in $G$, i.e., $l_{u,v}^*=l_{u,v}$.
\end{itemize}
These two assumptions are satisfied by many applications' data collection procedures.
For instance, the source data of online applications such as network traffic monitoring is given as a streaming of directed edges,
and the following simple method is computational and memory efficient for collecting edges and generating a small RESampled graph, which will be sent to remote network traffic analysis center:
Each incoming directed edge $u\to v$ is sampled when $\tau(u,v)\le \rho p$,
where $\rho$ is an integer (e.g., 10,000) and $\tau(u,v)$ is a hash function satisfying $\tau(u,v)=\tau(v,u)$ and mapping edges into integers $0,1,\ldots, \rho-1$ uniformly.
The property $\tau(u,v)=\tau(v,u)$ guarantees that edges $u\to v$ and $u\leftarrow v$ are sampled or discarded simultaneously.
Hence the label of a sampled edge $(u,v)\in E^*$ is the same as that of $(u,v)$ in $G$.
Using universal hashing~\cite{ChenKDD2006}, a simple instance of $\tau(u,v)$ is given as the following function when each $v \in V$ is an integer smaller than $\Delta$
\[
\tau(u,v)=(a (\min\{u,v\}\Delta  + \max\{u,v\} )+ b) \mod \gamma \mod \rho,
\]
where $\gamma$ is a prime larger than $\Delta^2$, $a$ and $b$ are any integers with $a\in \{1,\ldots, \rho-1\}$ and $b\in \{0,\ldots, \rho-1\}$.
We can easily find that $\tau(u,v)=\tau(v,u)$ and $\tau(u,v)$ maps edges into integers $0,1,\ldots, \rho-1$ uniformly.
The computational and space complexities of the above sampling method are both $O(1)$, which make it easy to use for data collections in practice.
\emph{As alluded before, in this paper, we aim to accurately
infer the motif concentrations of $G$ based on the given RESampled graph $G^*$.}

%!TEX root = samplingmotifs.tex
\section{Motif Statistical Inference} \label{sec:methods}
The motif statistics of RESampled graph $G^*$ and original graph $G$ can be quite different.
In this section, we introduce a probabilistic model to bridge the gap between the motif statistics of $G^*$ and $G$.
Using this model, we will show there exists
a simple and concise relationship between the motif statistics of $G$ and $G^*$.
We then  propose an efficient method
to infer the motif concentration of $G$ from $G^*$.
Finally, we also give a method to construct confidence intervals of our estimates of motif concentrations.
\subsection{Probabilistic Model of Motifs in $G^*$ and $G$}
To estimate the motif statistics of $G$ based on $G^*$,
we develop a probabilistic method to model the relationship between the motifs in $G^*$ and $G$.
Define $P=[P_{i,j}]$ where $P_{i,j}$ is the probability that $s^*$ is isomorphic to motif $M_i^{(k)}$
given that $s$ is isomorphic to motif $M_j^{(k)}$, i.e.,
%\[
$P_{i,j}=P(M(s^*)=M_i^{(k)} | M(s)=M_j^{(k)})$.
%\]

To obtain $P_{i,j}$, we first compute $\phi_{i,j}$,
which is the number of subgraphs of $M_j^{(k)}$ isomorphic to $M_i^{(k)}$.
For example, $M_2^{(3)}$ (i.e., the triangle) includes three subgraphs isomorphic to $M_1^{(3)}$ (i.e., the wedge)
for the undirected graph shown in Fig.~\ref{fig:345nodeclasses} (a).
Thus, we have $\phi_{1,2}=3$ for 3-node undirected motifs.
When $i=j$, $\phi_{i,j}=1$.
It is not easy to compute $\phi_{i,j}$ manually for 4- and 5-node motifs.
Hence we provide a simple method to compute $\phi_{i,j}$ in Algorithm~\ref{alg:tij}.
The computational complexity is $O(k^2k!)$.
Note that the cost of computing $\phi_{i,j}$ is not a big concern
%and can be neglected
because in practice, $k$ is usually five or less for motif discovery.
Denote by $V(s)$ and $E(s)$ the sets of nodes and edges in subgraph $s$ respectively.
We have the following equation
\[
P_{i,j}=\phi_{i,j} p^{|E(M_i^{(k)})|} q^{|E(M_j^{(k)})|-|E(M_i^{(k)})|},
\]
where $q=1-p$.
For all CISes in $G$ isomorphic to $M_j^{(k)}$, the above model tells us that
approximately $P_{i,j}\times 100\%$ of these CISes are expected to appear as CISes isomorphic to $M_i^{(k)}$ in $G^*$.

\begin{algorithm}[htb]
\caption{Pseudo-code of computing $\phi_{i,j}$, i.e., the number of subgraphs of $M_j^{(k)}$ that are isomorphic to $M_i^{(k)}$.}\label{alg:tij}
\begin{algorithmic}[1]
\State \textbf{Step 1}: Generate two graphs $\hat{G}=(\{v_1,\ldots,v_k\}, \hat{E}, \hat{L})$ and $\tilde{G}=(\{u_1,\ldots,u_k\}, \tilde{E}, \tilde{L})$,
                isomorphic to the motifs $M_i^{(k)}$ and $M_j^{(k)}$ respectively,
                where $\hat{E}$ and $\hat{L}$ are the edges and edge labels of nodes $\{v_1,\ldots,v_k\}$,
                and $\tilde{E}$ and $\tilde{L}$ are the edges and edge labels of nodes $\{u_1,\ldots,u_k\}$.
\State \textbf{Step 2}: Initialize a counter $y_{i,j}=0$. For each permutation $(x_1, \ldots, x_k)$ of integers $1, \ldots, k$,
                $y_{i,j}$ keeps unchanged when there exists an edge $(v_a, v_b)\in \hat{E}$ satisfying $(u_{x_a}, u_{x_b})\notin \tilde{E}$ or $\hat{l}_{v_a, v_b}\neq \tilde{l}_{u_{x_a}, u_{x_b}}$,
                and $y_{i,j}=y_{i,j}+1$ otherwise.
\State \textbf{Step 3}: Initialize a counter $z_j=0$. For each permutation $(x_1, \ldots, x_k)$ of integers $1, \ldots, k$,
                $z_j$ keeps unchanged when there exists an edge $(v_a, v_b)\in \hat{E}$ satisfying $(v_{x_a}, v_{x_b})\notin \hat{E}$ or $\hat{l}_{v_a, v_b}\neq \hat{l}_{v_{x_a}, v_{x_b}}$,
                and $z_j=z_j+1$ otherwise.
\State \textbf{Step 4}: Finally, $\phi_{i,j}=y_{i,j}/z_j$.
\end{algorithmic}
\end{algorithm}

\subsection{Motif Concentration Estimation}\label{subsec:ConcentrationEstimation}
Using the above probabilistic model,
we propose a method Minfer to estimate motif statistics of $G$ from $G^*$.
Denote by $m_i^{(k)}$, $1\le i \le T_k$, $k=3,4,\ldots$, the number of CISes in $G^*$ isomorphic to the motif $M_i^{(k)}$.
The method to compute $m_i^{(k)}$ is presented
in next section.
Then, the expectation of $m_i^{(k)}$ is computed as
\begin{equation}\label{eq: Emi}
\text{E}[m_i^{(k)}] = \sum_{1\le j\le T_k} n_j^{(k)} \, P_{i,j}.
\end{equation}
In matrix notation,
Equation ~(\ref{eq: Emi}) can be expressed as
\[
\text{E}[\mathbf{m}^{(k)}] = P \mathbf{n}^{(k)},
\]
where $P=[P_{ij}]_{1\le i,j\le T_k}$,
$\mathbf{n}^{(k)}=(n_1^{(k)}, \ldots, n_{T_k}^{(k)})\transpose$,
and $\mathbf{m}^{(k)}=(m_1^{(k)}, \ldots, m_{T_k}^{(k)})\transpose$.
Then, we have
\[
\mathbf{n}^{(k)} = P^{-1} \text{E}[\mathbf{m}^{(k)}].
\]
Thus, we estimate $\mathbf{n}^{(k)}$ as
\[
\mathbf{\hat n}^{(k)}= P^{-1} \mathbf{m}^{(k)},
\]
where $\mathbf{\hat n}^{(k)}=(\hat{n}_1^{(k)}, \ldots, \hat{n}_{T_k}^{(k)})\transpose$.
We easily have
\[
\text{E}[\mathbf{\hat n}^{(k)}]= \text{E}[P^{-1} \mathbf{m}^{(k)}]=P^{-1} \text{E}[\mathbf{m}^{(k)}] = \mathbf{n}^{(k)},
\]
therefore $\mathbf{n}^{(k)}$ is an unbiased estimator
of $\mathbf{n}^{(k)}$.
Finally, we estimate $\omega_i^{(k)}$ as follows
\begin{equation}\label{eq: estimator1}
\hat{\omega}_i^{(k)}=\frac{\hat{n}_i^{(k)}}{\sum_{j=1}^{T_k} \hat{n}_j^{(k)}}, \qquad 1\le i\le T_k.
\end{equation}
Denote by $\rho_i^{(k)}$ the concentration of motif $M_i^{(k)}$ in $G^*$, i.e.,  $\rho_i^{(k)}=\frac{m_i^{(k)}}{m^{(k)}}$.
Then we observe that ~(\ref{eq: estimator1}) is equivalent to the following equation, which directly describes the relationship between motif concentrations of $G$ and $G^*$.
Let $\bm{\hat{\omega}} = [\hat{\omega}_1^{(k)},\ldots, \hat{\omega}_{T_k}^{(k)}]\transpose$ and $\bm{\rho} = [\rho_1^{(k)},\ldots, \rho_{T_k}^{(k)}]\transpose$
\begin{equation}\label{eq: estimator2}
\bm{\hat{\omega}}=\frac{P^{-1} \bm{\rho}}{W},
\end{equation}
where $W=[1,\ldots,1] P^{-1} \bm{\rho}$ is a normalizer.
For 3-node undirected motifs, $P$ is computed as
\[
P=\left(
  \begin{array}{cc}
    p^2 & 3qp^2\\
    0 & p^3\\
  \end{array}
\right),
\]
and the inverse of $P$ is
\[
P^{-1}=\left(
  \begin{array}{cc}
    p^{-2} & -3qp^{-3}\\
    0 & p^{-3}\\
  \end{array}
\right).
\]

Expressions for $P$ and $P^{-1}$ for 3-node signed undirected motifs, 3-node directed motifs, 4-node undirected motifs, and 5-node undirected motifs can be found in Appendix.

\subsection{Lower Bound on Estimation Errors}

It is difficult to directly analyze the errors of our estimate $\bm{\hat{\omega}}$,
because it is complex to model the dependence of sampled CISes due to their shared edges and nodes.
Instead, we derive
a lower bound on the mean squared error (MSE) of $\bm{\hat{\omega}}$
using the Cram\'{e}r-Rao lower bound (CRLB) of $\bm{\hat{\omega}}$,
which gives the smallest MSE that any unbiased estimator of $\bm{\omega}$  can achieve.
For a $k$-node CIS $s$ selected from $k$-node CISes of $G$ at random,
the probability that $s$ is isomorphic to the $j$-th $k$-node motif is
$P(M(s)=M_j^{(k)})=\omega_j^{(k)}$.
Let $s^*$ be the induced subgraph of the node set $V(s)$ in the RESampled graph $G^*$.
Clearly, $s^*$ may not be connected. Furthermore there may exist nodes in $V(s)$ that are not present in $G^*$.
We say $s^*$ is \emph{evaporated} in $G^*$ for these two scenarios.
Let $P_{0,j}$ denote the probability that $s^*$ is evaporated given that its original CIS $s$ is isomorphic to the $j$-th $k$-node motif.
Then, we have
\[
P_{0,j}=1-\sum_{l=1}^{T_k} P_{l,j}.
\]
For a random $k$-node CIS $s$ of $G$, the probability that its associated $s^*$ in $G^*$ is isomorphic to the $i$-th $k$-node motif is
\[
\xi_i = P(M(s^*)=M_i^{(k)})=\sum_{j=1}^{T_k}  P_{i,j} \omega_j^{(k)}, \quad 1\le i\le T_k,
\]
and the probability that $s^*$ is evaporated is
\[
\xi_0 = \sum_{j=1}^{T_k} P_{0,j} \omega_j^{(k)}.
\]
When $s^*$ is evaporated, we denote $M(s^*)=0$.
Then, the likelihood function of $M(s^*)$ with respect to $\bm{\omega}^{(k)}$ is
\[
f(i| \bm{\omega}^{(k)})= \xi_i, \quad 0\le i\le T_k.
\]

The Fisher information of $M(s^*)$ with respect to $\bm{\omega}^{(k)}$ is defined as a matrix $J=[J_{i,j}]_{1\le i,j\le T_k}$, where
\begin{equation*}
\begin{split}
J_{i,j}&=\mathbb{E}\left[\frac{\partial \ln f(l| \bm{\omega}^{(k)})}{\partial \omega_i}\frac{\partial \ln f(l| \bm{\omega}^{(k)})}{\partial \omega_j}\right]\\
&=\sum_{l=0}^{T_k} \frac{\partial \ln f(l| \bm{\omega}^{(k)})}{\partial \omega_i}\frac{\partial \ln f(l| \bm{\omega}^{(k)})}{\partial \omega_j}\xi_l
= \sum_{l=0}^{T_k} \frac{P_{l,i} P_{l,j}}{\xi_l}.
\end{split}
\end{equation*}
For simplicity, we assume that the CISes of $G^*$ are independent (i.e., none overlapping edges). Then the Fisher information matrix of all $k$-node CISes is $n^{(k)} J$.
The Cram\'{e}r-Rao Theorem states that the MSE of any unbiased estimator is lower bounded by the inverse of the Fisher information matrix, i.e.,
\[
\text{MSE}(\hat{\omega}_i^{(k)})=\mathbb{E}[(\hat{\omega}_i^{(k)}-\omega_i^{(k)})^2]\ge \frac{(J^{-1})_{i,i}- \bm{\omega}^{(k)} (\bm{\omega}^{(k)})\transpose}{n^{(k)}}
\]
provided some weak regularity conditions hold~\cite{vanTrees}.
Here the term $\bm{\omega}^{(k)} (\bm{\omega}^{(k)})\transpose$ corresponds to the accuracy gain obtained by accounting for the constraint $\sum_{i=1}^{T_k} \omega_i^{(k)} = 1$.

\section{3-, 4-, and 5-node CIS Enumeration} \label{sec:enumerations}
The existing generalized graph enumeration method ~\cite{Wernicke2006} can be used for enumerating all $k$-node CISes in the RESampled graph $G^{*}$,
however it is complex to apply and is inefficient for small values of $k= 3, 4, 5$.
In this section, we first present a method (an extension of the NodeIterator++ method in~\cite{SuriWWW2011}) to enumerate all 3-node CISes in $G^{*}$.
Then, we propose new methods to enumerate 4 and 5-node CISes in $G^{*}$ respectively.
In what follows we denote $N^*(u)$ as the neighbors of $u$ in $G^*$.
Note that in this section $G^{*}$ is the default graph when a function's underlying graph is omitted for simplicity.
For example, the CIS with nodes $u$, $v$, and $w$ refers to the CIS with nodes $u$, $v$, and $w$ in $G^*$.

\subsection{3-node CIS Enumeration}
Algorithm~\ref{alg:3nodeCISEnumeration} shows our 3-node CISes enumeration method.
Similar to the NodeIterator++ method in~\cite{SuriWWW2011}, we ``pivot" (the associated operation is discussed later) each node $u\in V^*$ to enumerate CISes including $u$.
For any two neighbors $v$ and $w$ of $u$, we can easily find that the induced graph $s$ with nodes $u$, $v$ and $w$ is a 3-node CIS.
Thus, we enumerate all pairs of two nodes in $N^*(u)$, and update their associated 3-node CIS for $u$.
We call this process ``pivoting" $u$ for 3-node CISes.
\begin{algorithm}
\SetKwFunction{continue}{continue}
\SetKwFunction{induced}{induced}
\SetKwInOut{Input}{input}
\SetKwInOut{Output}{output}
\Input{$G^*=(V^*, E^*, L^*)$}
\tcc{\footnotesize $m_i^{(3)}$ records the number of CISes in $G^*$  isomorphic to motif $M_i^{(3)}$, $1 \le i \le T_3$.}
\Output{$\mathbf{m}^{(3)}=(m_1^{(3)}, \ldots, m_{T_3}^{(3)})\transpose$}
\BlankLine
\For {$u \in V^*$} {
    \For {$v \in N^*(u)$} {
        \For {$w \in N^*(u)$ and $w \succ v$} {
           \tcc{\footnotesize $\induced(G^*, \Gamma)$ returns the CIS with the node set $\Gamma$ of $G^*$.}
            $s \gets\induced(G^*, \{u, v, w\})$\;
            \If {$(w,v)\in E^*$ and $u\succ v$} {
                $\continue()$;
            }
            \tcc{$M(s)$ is the motif class ID of $s$.}
            $i \gets M(s)$\;
            $m_i^{(3)} \gets m_i^{(3)} + 1$\;
        }
    }
}
\caption{3-node CIS enumeration. \label{alg:3nodeCISEnumeration}}
\end{algorithm}

Clearly, a 3-node CIS $s$ is counted three times when the associated undirected graph of $s$ by discarding edge
labels is isomorphic to a triangle, once by pivoting each node $u$, $v$, and $w$.
Let $\succ$ be a total order on all of the nodes,
which can be easily defined and obtained, e.g. from array position or pointer addresses.
To ensure each CIS is enumerated once and only once,
we let one and only one node in each CIS be ``responsible" for making sure the CIS gets counted.
When we ``pivot" $u$ and enumerate a CIS $s$, $s$ is counted if $u$ is the `responsible" node of $s$.
Otherwise, $s$ is discarded and not counted.
We use the same method in~\cite{Schank2007,SuriWWW2011}, i.e., let the node with lowest order in a CIS whose associated undirected graph isomorphic to a triangle be the ``responsible" node.
For the other classes of CISes, their associated undirected graphs are isomorphic to an unclosed wedge, i.e., the first motif in Fig.~\ref{fig:345nodeclasses} (a).
For each of these CISes, we let the node in the middle of its associated undirected graph (e.g., the node with degree 2 in the unclosed wedge) be the ``responsible" node.

\subsection{4-node CIS Enumeration}
Algorithm~\ref{alg:4nodeCISEnumeration} shows our 4-node CISes enumeration method.
To enumerate 4-node CISes, we ``pivoting" each node $u$ as follows:
For each pair of $u$'s neighbors $v$ and $w$ where $w\succ v$,
we compute the neighborhood of $u$, $v$, and $w$ , defined as $\Gamma = N^*(u) \cup N^*(v) \cup N^*(w) - \{u, v, w\}$.
For any node $x\in \Gamma$, we observe that the induced graph $s$ consisting of nodes $u$, $v$, $w$, and $x$ is a 4-node CIS.
Thus, we enumerate each node $x$ in $\Gamma$, and update the 4-node CIS consisting of $u$, $v$, $w$, and $x$.
We repeat this process until all pairs of $u$'s neighbors $v$ and $w$ are enumerated and processed.

\begin{figure}[htb]
\begin{center}
\includegraphics[width=0.28\textwidth]{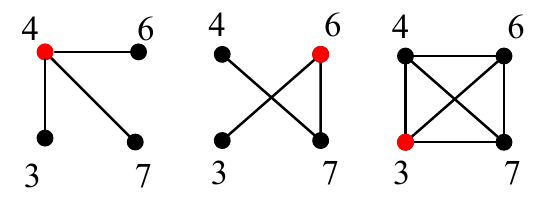}
\caption{Examples of ``responsible" nodes of 4-node CISs.
Graphs shown are CISes' associated undirected graphs, and the number near to a node represents the node order.
Red nodes are ``responsible" nodes.}
\label{fig:response4node}
\end{center}
\end{figure}

Similar to 3-node CISes, some 4-node CISes might be enumerated and counted more than once when we ``pivoting" each node $u$ as above.
To solve this problem, we propose the following methods for making sure each 4-node CIS $s$ is enumerated and gets counted once and only once:
When $(u,x)\in E^*$ and $w\succ x$, we discard $x$.
Otherwise, denote by $\hat s$ the associated undirected graph of $s$ by discarding edge labels.
When $\hat s$ includes one and only one node $u$ having at least 2 neighbors in $V(\hat s)$,
we let $u$ be the ``responsible" of $s$.
For example, the node 4 is the ``responsible" node of the first subgraph in Fig.~\ref{fig:response4node}.
When $\hat s$ includes more than one node having at least 2 neighbors in $V(\hat s)$,
we let the node with lowest order among the nodes having at least 2 neighbors in $V(\hat s)$ be the ``responsible" node of $s$.
For example, the nodes 6 and 3 are the ``responsible" nodes of the second and third subgraphs in Fig.~\ref{fig:response4node}.

\begin{algorithm}[htb]
\SetKwFunction{findNodes}{findNodes}
\SetKwFunction{minNodes}{minNodes}
\SetKwFunction{induced}{induced}
\SetKwFunction{undirected}{undirected}
\SetKwFunction{continue}{continue}
\SetKwInOut{Input}{input}
\SetKwInOut{Output}{output}
\Input{$G^*=(V^*, E^*, L^*)$}
\tcc{\footnotesize $m_i^{(4)}$ records the number of CISes in $G^*$  isomorphic to motif $M_i^{(4)}$, $1 \le i \le T_4$.}
\Output{$\mathbf{m}^{(4)}=(m_1^{(4)}, \ldots, m_{T_4}^{(4)})\transpose$}
\BlankLine
\For {$u \in V^*$} {
    \For {$v \in N^*(u)$} {
        \For {$w \in N^*(u)$ and $w \succ v$} {
            $\Gamma = N^*(u) \cup N^*(v) \cup N^*(w) - \{u, v, w\}$\;
            \For {$x \in \Gamma$} {
                \If {$(u, x)\in E^*$ and $w \succ x$} {
                    $\continue()$;
                }
               \tcc{\footnotesize $\induced(G^*, \{u, v, w, x\})$ is defined same as Alg.~\ref{alg:3nodeCISEnumeration}.}
                $s \gets\induced(G^*, \{u, v, w, x\})$\;
                \tcc{\footnotesize $\undirected(s)$ returns the associated undirected graph of $s$ by discarding edge labels.}
                $\hat s \gets \undirected(s)$\;
                \tcc{\footnotesize $\findNodes(\hat s, t)$ returns the set of nodes in $V(\hat s)$ having at least $t$ neighbors in $V(\hat s)$.}
                $\Lambda \gets \findNodes(\hat s, 2)$\;
                \If {$|\Lambda| \ge 2$} {
                    \tcc{\footnotesize $\minNodes(\Lambda)$ returns the node with the lowest order in $V(\hat s)$.}
                    \If {$u \succ \minNodes(\Lambda)$} {
                        $\continue()$;
                    }
                }
                $i \gets M(s)$\;
                $m_i^{(4)} \gets m_i^{(4)} + 1$\;
            }
        }
    }
}
\caption{4-node CIS enumeration. \label{alg:4nodeCISEnumeration}}
\end{algorithm}

\subsection{5-node CIS Enumeration}
Algorithm~\ref{alg:5nodeCISEnumeration} describes our 5-node CISes enumeration method.
For a 5-node CIS $s$, we classify it into two types according to its associated undirected graph $\hat s$:
\begin{itemize}
  \item \textbf{5-node CIS $s$ with type 1}: $\hat s$ has at least one node having more than two neighbors in $V(\hat s)$;
  \item \textbf{5-node CIS $s$ with type 2}: $\hat s$ has no node having more than two neighbors in $V(\hat s)$, i.e., $\hat s$ is isomorphic to a 5-node line or a circle, i.e., the first or sixth motifs in Fig.~\ref{fig:345nodeclasses} (c).
\end{itemize}
We propose two different methods to enumerate these two types of 5-node CISes respectively.

\begin{figure}[htb]
\begin{center}
\includegraphics[width=0.46\textwidth]{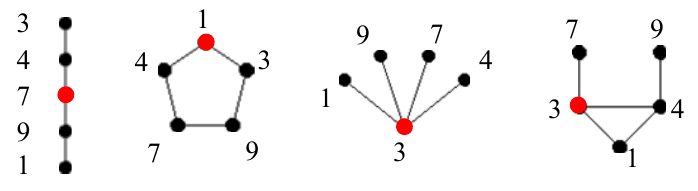}
\caption{Examples of ``responsible" nodes of 5-node CISs.
Graphs shown are CISes' associated undirected graphs, and the number near to a node represents the node order.
Red nodes are ``responsible" nodes.}
\label{fig:response5node}
\end{center}
\end{figure}

To enumerate 5-node CISes with type 1, we ``pivoting" each node $u$ as follows:
When $u$ has at least three neighbors,
we enumerate each combination of three nodes $v, w, x\in N^*(u)$ where $x\succ w\succ v$, and then compute the neighborhood of $u$, $v$, $w$, and $x$, defined as $\Gamma \gets N^*(u) \cup N^*(v) \cup N^*(w) \cup N^*(x) - \{u, v, w, x\}$.
For any node $y\in \Gamma$, we observe that the induced graph $s$ consisting of nodes $u$, $v$, $w$, $x$, and $y$ is a 5-node CIS.
Thus, we enumerate each node $y$ in $\Gamma$, and update the associated 5-node CIS consisting of $u$, $v$, $w$, $x$, and $y$.
We repeat this process until all combinations of three nodes $v, w, x\in N^*(u)$ are enumerated and processed.
Similar to 4-node CISes, we propose the following method to make sure each 5-node $s$ is enumerated and gets counted once and only once:
When $(y, u)\in E^*$ and $y\succ x$, we discard $y$.
Otherwise, let $\hat s$ be the associated undirected graph of $s$, and we then pick the node with lowest order among the nodes having more than two neighbors in $V(\hat s)$ be the ``responsible" node.
The third and fourth subgraphs in Fig.~\ref{fig:response5node} are two corresponding examples.

To enumerate 5-node CISes with type 2, we ``pivoting" each node $u$ as follows:
When $u$ has at least two neighbors,
we first enumerate each pair of $u$'s neighbors $v$ and $w$ where $(v,w)\notin E^*$.
Then, we compute $\Gamma_v$ defined as the set of $v$'s neighbors not including $u$ and $w$ and not connecting to $u$ and $w$,
that is, $\Gamma_v \gets N^*(v) - \{u, w\} - N^*(u) - N^*(w)$.
Similarly, we compute $\Gamma_w$ defined as the set of $w$'s neighbors not including $u$ and $v$ and not connecting to $u$ and $v$,
i.e., $\Gamma_w \gets N^*(w) - \{u, v\} - N^*(u) - N^*(v)$.
Clearly, $\Gamma_v\cap \Gamma_w=\varnothing$.
For any $x\in \Gamma_v$ and $y\in \Gamma_w$,
we observe that the induced graph $s$ consisting of nodes $u$, $v$, $w$, $x$, and $y$ is a 5-node CIS with type 2.
Thus, we enumerate each pair of $x\in \Gamma_v$ and $y\in \Gamma_w$, and update the 5-node CIS consisting of $u$, $v$, $w$, $x$, and $y$.
We repeat this process until all pairs of $u$'s neighbors $v$ and $w$ are enumerated and processed.
To make sure each CIS $s$ is enumerated and gets counted once and only once,
we let the node with lowest order be the ``responsible" node when the associated undirected graph $\hat s$ of $s$ isomorphic to a 5-node circle.
When $\hat s$ is isomorphic to a 5-node line,
we let the node in the middle of the line be the ``responsible" node.
The first and second subgraphs in Fig.~\ref{fig:response5node} are two examples respectively.

\begin{algorithm}
\SetKwFunction{findNodes}{findNodes}
\SetKwFunction{minNodes}{minNodes}
\SetKwFunction{induced}{induced}
\SetKwFunction{undirected}{undirected}
\SetKwFunction{continue}{continue}
\SetKwInOut{Input}{input}
\SetKwInOut{Output}{output}
\Input{$G^*=(V^*, E^*, L^*)$}
\tcc{\footnotesize $m_i^{(5)}$ records the number of CISes in $G^*$  isomorphic to motif $M_i^{(5)}$, $1 \le i \le T_5$.}
\Output{$\mathbf{m}^{(5)}=(m_1^{(5)}, \ldots, m_{T_5}^{(5)})\transpose$}
\BlankLine
\tcc{\footnotesize The functions $\findNodes$, $\minNodes$, $\induced$, and $\undirected$ are defined in Algorithms~\ref{alg:3nodeCISEnumeration} and~\ref{alg:4nodeCISEnumeration}.}
\For {$u \in V^*$} {
    \For {$v \in N^*(u)$} {
        \For {$w \in N^*(u)$ and $w \succ v$} {
            \tcc{\footnotesize Enumerate and update CIS $s$ with $\undirected(s)$ not isomorphic to a 5-node line and circle.}
            \For {$x \in N^*(u)$ and $x \succ w$} {
                $\Gamma \gets N^*(u) \cup N^*(v) \cup N^*(w) \cup N^*(x) - \{u, v, w, x\}$\;
                \For {$y \in \Gamma$} {
                    \If {$(y, u)\in E^*$ and $x \succ y$} {
                        $\continue()$;
                    }
                    $s \gets\induced(G^*, \{u, v, w, x, y\})$\;
                    $\hat s \gets \undirected(s)$\;
                    $\Lambda \gets \findNodes(\hat s, 3)$\;
                    \If {$|\Lambda| \ge 2$} {
                        \If {$u \succ \minNodes(\Lambda)$} {
                            $\continue()$;
                        }
                    }
                    $i \gets M(s)$\;
                    $m_i^{(5)} \gets m_i^{(5)} + 1$\;
                }
            }

            \tcc{\footnotesize Enumerate and update $s$ with $\undirected(s)$ isomorphic to a 5-node line or circle.}
            \If {$(u, v)\notin E^*$} {
                $\Gamma_v \gets N^*(v) - \{u, w\} - N^*(u) - N^*(w)$\;
                \For {$x \in \Gamma_v$} {
                    \tcc{\footnotesize  $s$ with $\undirected(s)$ isomorphic to a 5-node circle.}
                    $\Gamma_w \gets N^*(w) - \{u, v\} - N^*(u) - N^*(v)$\;
                    \For {$y \in \Gamma_w$} {
                        \If {$(x,y)\in E^*$ and $u \succ \minNodes(\{u, v, w, x, y\})$} {
                            $\continue()$;
                        }
                        $s \gets\induced(G^*, \{u, v, w, x, y\})$\;
                        $i \gets M(s)$\;
                        $m_i^{(5)} \gets m_i^{(5)} + 1$\;
                    }
                }
            }
        }
    }
}
\caption{5-node CIS enumeration. \label{alg:5nodeCISEnumeration}}
\end{algorithm}

%!TEX root = samplingmotifs.tex
\begin{table}[htb]
\begin{center}
\caption{Graph datasets used in our simulations, ``edges" refers to the number of edges in the
undirected graph generated by discarding edge labels, ``max-degree" represents  the maximum number of edges incident to a node in the undirected graph.\label{tab:datasets}}
\begin{tabular}{|c|c|c|c|}
\hline
{\bf Graph}&{\bf nodes}& {\bf edges}& {\bf max-degree}\\
\hline \hline
Flickr~\cite{MisloveIMC2007}&1,715,255&15,555,041&27,236\\
Pokec~\cite{Takac2012}&1,632,803&22,301,964&14,854\\
LiveJournal~\cite{MisloveIMC2007}&5,189,809&48,688,097&15,017\\
YouTube~\cite{MisloveIMC2007}&1,138,499&2,990,443&28,754\\
Wiki-Talk~\cite{Leskovec2010}&2,394,385&4,659,565&100,029\\
Web-Google~\cite{GoogleProgrammingContest2002}&875,713&	4,322,051&6,332\\
soc-Epinions1~\cite{Richardson2003}&75,897&405,740&3,044\\
soc-Slashdot08~\cite{LeskovecIM2009}&77,360&469,180&2,539\\
soc-Slashdot09~\cite{LeskovecIM2009}&82,168&504,230&2,552\\
sign-Epinions~\cite{LeskovecCHI2010}&119,130&704,267&3,558\\
sign-Slashdot08~\cite{LeskovecCHI2010}&77,350&416,695&2,537\\
sign-Slashdot09~\cite{LeskovecCHI2010}&82,144&504,230&2,552\\
com-DBLP~\cite{YangICDM2012}&317,080&1,049,866&343\\
com-Amazon~\cite{YangICDM2012}&334,863&925,872&549\\
p2p-Gnutella08~\cite{Ripeanu2002}&6,301&20,777&97\\
ca-GrQc~\cite{LeskovecTKDD2007}&5,241&14,484&81\\
ca-CondMat~\cite{LeskovecTKDD2007}&23,133&93,439&279\\
ca-HepTh~\cite{LeskovecTKDD2007}&9,875&25,937&65\\
\hline
\end{tabular}
\end{center}
\end{table}

\begin{table}[htb]
\begin{center}
\caption{Values of $\omega_i^{(3)}$, the concentrations of 3-node undirected and directed motifs.
Flickr, Pokec, LiveJournal,Wiki-Talk, and Web-Google have
$1.35\times 10^{10}$, $2.02\times 10^9$, $6.90\times 10^9$, $1.2\times 10^{10}$, and $7.00\times 10^8$ 3-node CISes respectively.
($i$ is the motif ID.)
\label{tab:3nodeundergroundth}}
\begin{tabular}{|c|ccccc|}
\hline
\multirow{2}{*}{$i$}&\multirow{2}{*}{Flickr}&\multirow{2}{*}{Pokec}&LiveLive-&Wiki-&Web-\\
&&&Journal&Talk&Google\\
\hline
\hline
\multicolumn{6}{|c|}{undirected 3-node motifs}\\
\hline
1&9.60e-01&9.84e-01&9.55e-01&9.99e-01&9.81e-01\\
2&4.04e-02&1.60e-02&4.50e-02&7.18e-04&1.91e-02\\
\hline
\hline
\multicolumn{6}{|c|}{directed 3-node motifs}\\
\hline
1&2.17e-01&1.77e-01&7.62e-02&8.91e-01&1.27e-02\\
2&6.04e-02&1.11e-01&4.83e-02&4.04e-02&1.60e-02\\
3&1.28e-01&1.60e-01&3.28e-01&3.91e-03&9.28e-01\\
4&2.44e-01&1.74e-01&1.14e-01&5.43e-02&3.09e-03\\
5&1.31e-01&1.91e-01&1.73e-01&5.48e-03&1.92e-02\\
6&1.80e-01&1.71e-01&2.15e-01&3.88e-03&1.92e-03\\
7&5.69e-05&7.06e-05&2.74e-05&1.37e-05&4.91e-05\\
8&6.52e-03&2.49e-03&8.66e-03&1.81e-04&6.82e-03\\
9&1.58e-03&1.03e-03&1.06e-03&8.42e-05&2.84e-04\\
10&5.19e-03&1.91e-03&6.63e-03&1.28e-04&2.77e-03\\
11&6.46e-03&2.03e-03&6.27e-03&8.03e-05&5.98e-03\\
12&1.07e-02&5.13e-03&9.82e-03&1.78e-04&1.21e-03\\
13&9.86e-03&3.45e-03&1.26e-02&6.65e-05&2.00e-03\\
\hline
\end{tabular}
\end{center}
\end{table}

\begin{table}[htb]
\begin{center}
\caption{NRMSEs of $\hat\omega_i^{(3)}$, the concentration estimates of 3-node undirected motifs for $p=0.01$ and $p=0.05$ respectively.
($i$ is the motif ID.)
\label{tab:results3undirected}}
\begin{tabular}{|c|ccccc|}
\hline
\multirow{2}{*}{$i$}&\multirow{2}{*}{Flickr}&\multirow{2}{*}{Pokec}&LiveLive-&Wiki-&Web-\\
&&&Journal&Talk&Google\\
\hline
\hline
\multicolumn{6}{|c|}{$p=0.01$}\\
\hline
1&1.92e-03&3.26e-03&2.69e-03&5.21e-03&2.93e-04\\
2&4.56e-02&6.92e-02&1.64e-01&2.67e-01&4.00e-01\\
\hline
\multicolumn{6}{|c|}{$p=0.05$}\\
\hline
1&2.90e-04&4.10e-04&2.64e-04&6.06e-04&2.92e-05\\
2&6.90e-03&8.68e-03&1.61e-02&3.11e-02&3.99e-02\\
\hline
\end{tabular}
\end{center}
\end{table}

\begin{table}[htb]
\begin{center}
\caption{Values of $\omega_i^{(3)}$, the concentrations of 3-node signed and undirected motifs.
Sign-Epinions, sign-Slashdot08, sign-Slashdot09 have $1.72\times 10^8$,  $6.72\times 10^7$, and $7.25\times 10^7$ 3-node CISes respectively.
($i$ is the motif ID.)
\label{tab:3nodesignedgroundth}}
\begin{tabular}{|c|ccc|}
\hline
$i$&sign-Epinions&sign-Slashdot08&sign-Slashdot09\\
\hline
1&6.69e-01&6.58e-01&6.68e-01\\
2&2.12e-01&2.32e-01&2.25e-01\\
3&9.09e-02&1.02e-01&9.96e-02\\
4&2.29e-02&5.86e-03&5.75e-03\\
5&2.76e-03&9.74e-04&9.34e-04\\
6&2.49e-03&1.14e-03&1.13e-03\\
7&3.81e-04&1.80e-04&1.76e-04\\
\hline
\end{tabular}
\end{center}
\end{table}

\begin{table}[htb]
\begin{center}
\caption{Values of $\omega_i^{(4)}$, the concentrations of 4-node undirected motifs.
Soc-Epinions1, soc-Slashdot08, soc-Slashdot09, and com-Amazon have $2.58\times 10^{10}$, $2.17\times 10^{10}$, $2.42\times 10^{10}$, and $1.78\times 10^8$ 4-node CISes respectively.
($i$ is the motif ID.)
\label{tab:4nodeundergroundth}}
\begin{tabular}{|c|cccc|}
\hline
\multirow{2}{*}{$i$}&soc-&soc-&soc-&com-\\
&Epinions1&Slashdot08&Slashdot09&Amazon\\
\hline
1&3.24e-01&2.93e-01&2.90e-01&2.10e-01\\
2&6.15e-01&6.86e-01&6.89e-01&6.99e-01\\
3&2.78e-03&1.25e-03&1.30e-03&2.37e-03\\
4&5.45e-02&1.86e-02&1.84e-02&7.69e-02\\
5&3.01e-03&7.77e-04&8.48e-04&1.05e-02\\
6&2.25e-04&9.19e-05&9.36e-05&1.55e-03\\
\hline
\end{tabular}
\end{center}
\end{table}

\begin{table}[htb]
\begin{center}
\caption{Values of $\omega_i^{(5)}$, concentrations of 5-node undirected motifs.
Com-Amazon, com-DBLP, p2p-Gnutella08, ca-GrQc, ca-CondMat, and ca-HepTh have
$8.50\times 10^9$, $3.34\times 10^{10}$, $3.92\times 10^8$, $3.64\times 10^7$, $3.32\times 10^9$, and $8.73\times 10^7$ 5-node CISes respectively.
($i$ is the motif ID.)
\label{tab:5nodeundergroundth}}
\begin{tabular}{|c|cccccc|}
\hline
\multirow{2}{*}{$i$}&com-A&com-&p2p-Gn&ca-&ca-Con&ca-\\
&mazon&DBLP&utella08&GrQc&dMat&HepTh\\
\hline
1&2.9e-2&1.4e-1&2.6e-1&9.8e-2&1.4e-1&2.6e-1\\
2&7.5e-1&1.8e-1&1.8e-1&5.2e-2&2.2e-1&8.2e-2\\
3&1.6e-1&4.4e-1&4.6e-1&2.1e-1&4.3e-1&4.4e-1\\
4&6.0e-3&4.8e-2&1.1e-2&1.0e-1&4.9e-2&6.0e-2\\
5&2.3e-3&1.1e-3&2.7e-2&1.4e-3&2.1e-3&5.4e-3\\
6&3.6e-5&5.0e-5&1.4e-3&9.2e-5&1.1e-4&4.1e-4\\
7&1.5e-2&5.6e-2&2.7e-2&1.1e-1&5.5e-2&6.4e-2\\
8&3.5e-2&7.9e-2&2.2e-2&1.2e-1&8.0e-2&5.2e-2\\
9&1.4e-3&4.2e-3&1.4e-3&1.5e-2&7.0e-3&8.4e-3\\
10&1.7e-4&1.4e-4&1.0e-3&6.5e-4&3.0e-4&8.0e-4\\
11&7.3e-3&8.1e-3&4.3e-3&2.3e-2&9.9e-3&1.0e-2\\
12&5.3e-4&6.4e-3&2.8e-4&2.3e-2&4.5e-3&3.6e-3\\
13&8.2e-5&3.5e-6&7.4e-4&4.5e-6&6.4e-6&3.5e-5\\
14&3.9e-4&5.2e-4&1.7e-4&2.8e-3&6.6e-4&1.0e-3\\
15&6.7e-4&2.6e-2&7.6e-5&1.5e-1&5.9e-3&5.3e-3\\
16&7.1e-4&3.4e-4&1.4e-4&1.4e-3&9.2e-4&4.4e-4\\
17&3.9e-5&1.1e-5&8.0e-5&4.3e-5&2.9e-5&8.4e-5\\
18&2.3e-5&4.9e-6&6.0e-6&2.3e-5&8.5e-6&3.0e-5\\
19&2.4e-4&2.8e-3&1.5e-5&1.9e-2&9.8e-4&5.8e-4\\
20&5.8e-5&4.2e-4&7.0e-7&8.0e-3&1.4e-4&8.2e-5\\
21&7.2e-6&7.9e-3&1.5e-8&6.1e-2&1.5e-4&3.2e-3\\
\hline
\end{tabular}
\end{center}
\end{table}

\section{Evaluation} \label{sec:results}
In this section, we first introduce our experimental datasets.
Then we present results of experiments used to evaluate the performance of our method, Minfer, for characterizing CIS classes of size $k=3, 4, 5$.
\subsection{Datasets}
We evaluate the performance of our methods on publicly available datasets taken from the Stanford Network Analysis Platform (SNAP)\footnote{www.snap.stanford.edu}, which are summarized in Table~\ref{tab:datasets}.
We start by evaluating the performance of our methods in characterizing $3$-node CISes over million-node graphs: Flickr, Pokec, LiveJournal, YouTube, Web-Google, and Wiki-talk, contrasting our results with the ground truth computed through an exhaustive method.
It is computationally intensive to calculate the ground-truth of $4$-node and $5$-nodes CIS classes in large graphs.
For example, we can easily observe that a node with degree $d>4$ is included in at least $\frac{1}{6}d(d-1)(d-2)$ 4-node CISes and $\frac{1}{24}d(d-1)(d-2)(d-3)$ 5-node CISes, therefore it requires more than $O(10^{15})$ and $O(10^{19})$ operations to enumerate the 4-node and 5-node CISes of the Wiki-talk graph, which has a node with 100,029 neighbors.
Even for a relatively small graph such as soc-Slashdot08, it takes almost 20 hours to compute all of its 4-node CISes.
To solve this problem, the experiments for 4-node CISes are performed on four medium-sized graphs soc-Epinions1, soc-Slashdot08, soc-Slashdot09, com-DBLP, and com-Amazon,
and the experiments for 5-node CISes are performed on four relatively small graphs ca-GR-QC, ca-HEP-TH, ca-CondMat, and p2p-Gnutella08,
where computing the ground-truth is feasible.
We also evaluate the performance of our methods for characterizing signed CIS classes in graphs sign-Epinions, sign-Slashdot08, and sign-Slashdot09.

\subsection{Error Metric}
In our experiments, we focus on the normalized root mean square error (NRMSE) to measure the relative error of the estimator $\hat{\omega}_i$ of the subgraph class concentration $\omega_i$, $i=1,2,\dots$.
$\text{NRMSE}(\hat{\omega}_i)$ is defined as:
\[
\text{NRMSE}(\hat{\omega}_i)=\frac{\sqrt{\text{MSE}(\hat{\omega}_i)}}{\omega_i}, \qquad i=1,2,\dots,
\]
where $\text{MSE}(\hat{\omega}_i)$ is defined as
the mean square error (MSE) of an estimate $\hat{\omega}$ with respect to its true value $\omega>0$,
that is
\[
\text{MSE}(\hat{\omega})=\mathbb{E}[(\hat{\omega}-\omega)^2]=\text{var}(\hat{\omega})+\left(\mathbb{E}[\hat{\omega}]-\omega\right)^2.
\]
We note that $\text{MSE}(\hat{\omega})$ decomposes into a sum of the variance and bias of the estimator $\hat{\omega}$.
Both quantities are important and need to be as small as possible to achieve good estimation performance.
When $\hat{\omega}$ is an unbiased estimator of $\omega$,
then we have $\text{MSE}(\hat{\omega})= \text{var}(\hat{\omega})$
and thus $\text{NRMSE}(\hat{\omega}_i)$ is equivalent to the normalized standard error of $\hat{\omega}_i$, i.e., $\text{NRMSE}(\hat{\omega}_i)= \sqrt{\text{var}(\hat{\omega}_i)}/\omega_i$.
Note that our metric uses
the relative error. Thus, when $\omega_i$ is small, we consider
values as large as $\text{NRMSE}(\hat{\omega}_i)=1$ to be acceptable.
In all our experiments, we average the estimates and calculate their
NRMSEs over 1,000 runs.

\begin{figure*}[htb]
\center
\subfigure[$p=0.01$]{
\includegraphics[width=0.9\textwidth]{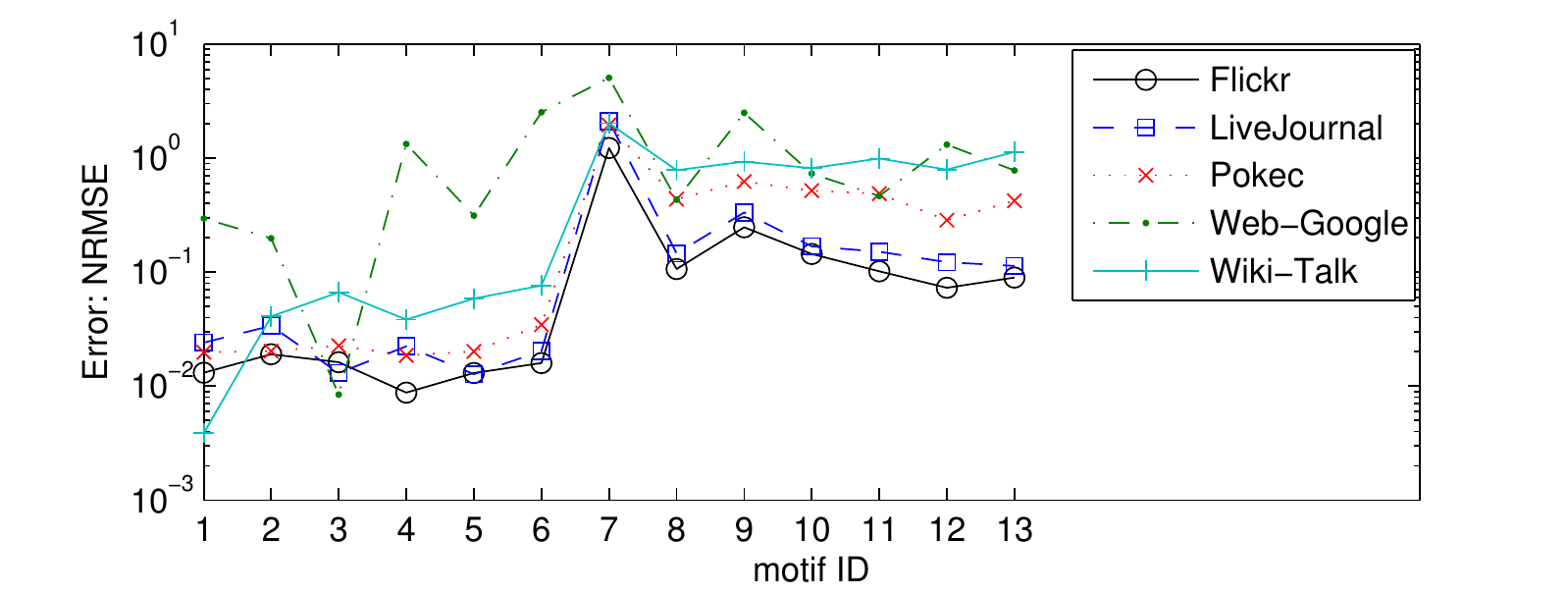}}
\subfigure[$p=0.05$]{
\includegraphics[width=0.9\textwidth]{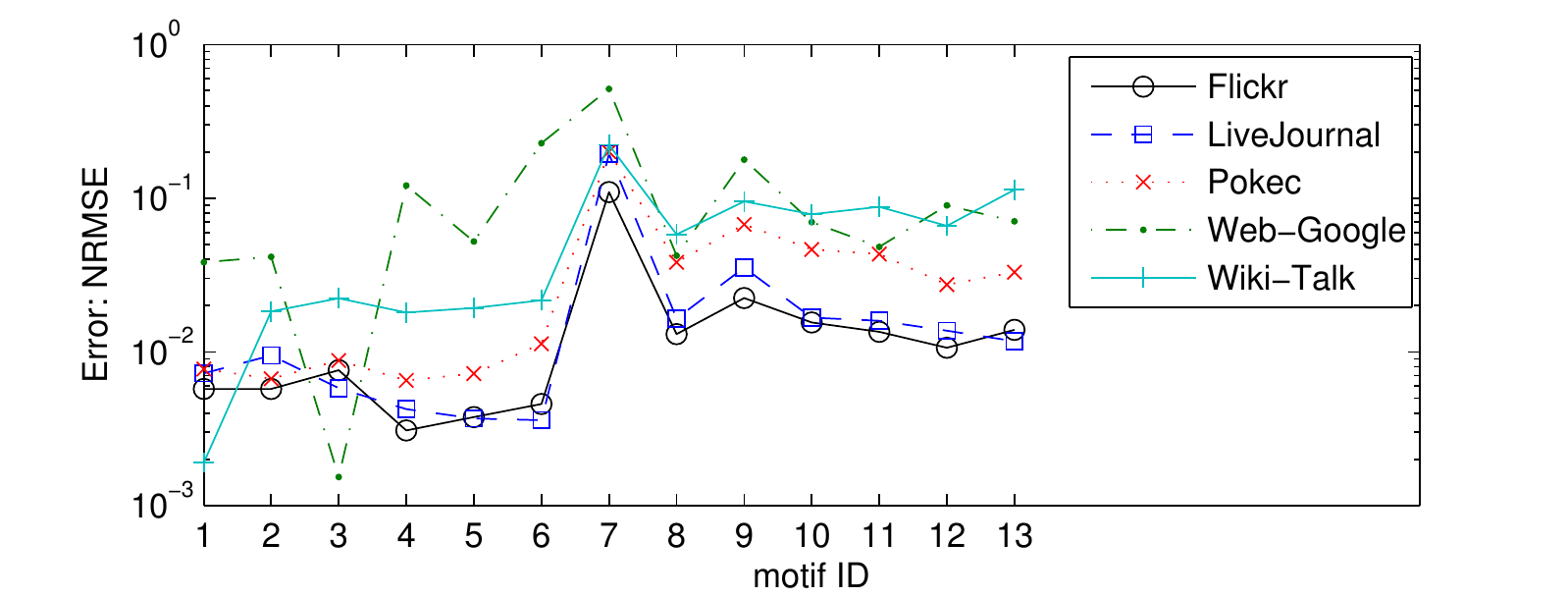}}
\caption{NRMSEs of $\hat\omega_i^{(3)}$, the concentration estimates of 3-node directed motifs for $p=0.01$ and $p=0.05$ respectively.}
\label{fig:Results3nodeDirectedCIS}
\end{figure*}

\subsection{Accuracy Results}
\subsubsection{Accuracy of inferring 3-node motifs' concentrations}
Table~\ref{tab:3nodeundergroundth} shows the real values of the 3-node undirected and directed motifs' concentrations for the undirected graphs and directed graphs of Flickr, Pokec, LiveJournal,Wiki-Talk, and Web-Google.
Among all 3-node directed motifs, the 7-th motif exhibits the smallest concentration for all these five directed graphs.
Here the undirected graphs are obtained by discarding the edge directions of directed graphs.
Flickr, Pokec, LiveJournal,Wiki-Talk, and Web-Google have $1.35\times 10^{10}$, $2.02\times 10^9$, $6.90\times 10^9$, $1.2\times 10^{10}$, and $7.00\times 10^8$ 3-node CISes respectively.
Table~\ref{tab:results3undirected} shows the NRMSEs of our estimates of 3-node undirected motifs' concentrations for $p=0.01$ and $p=0.05$ respectively.
We observe that the NRMSEs associated with the sampling probability $p=0.05$ is about ten times smaller than the NRMSEs when $p=0.01$.
The NRMSEs are smaller than 0.04 when $p=0.05$ for all five graphs.
Fig.~\ref{fig:Results3nodeDirectedCIS} shows the NRMSEs of our estimates of 3-node directed motifs' concentrations for $p=0.01$ and $p=0.05$ respectively.
Similarly, we observe the NRMSEs when $p=0.05$ are nearly ten times smaller than the NRMSEs when $p=0.01$.
The NRMSE of our estimates of $\omega_7^{(3)}$ (i.e., the 7-th 3-node directed motif concentration) exhibits the largest error.
Except for $\omega_7^{(3)}$, the NRMSEs of the other motif concentrations' estimates are smaller than 0.2 when $p=0.05$.

\begin{figure}[htb]
\center
%\subfigure[$p=0.01$]{
%\includegraphics[width=0.225\textwidth]{3nodeSignedCISP01}}
\subfigure[$p=0.05$]{
\includegraphics[width=0.225\textwidth]{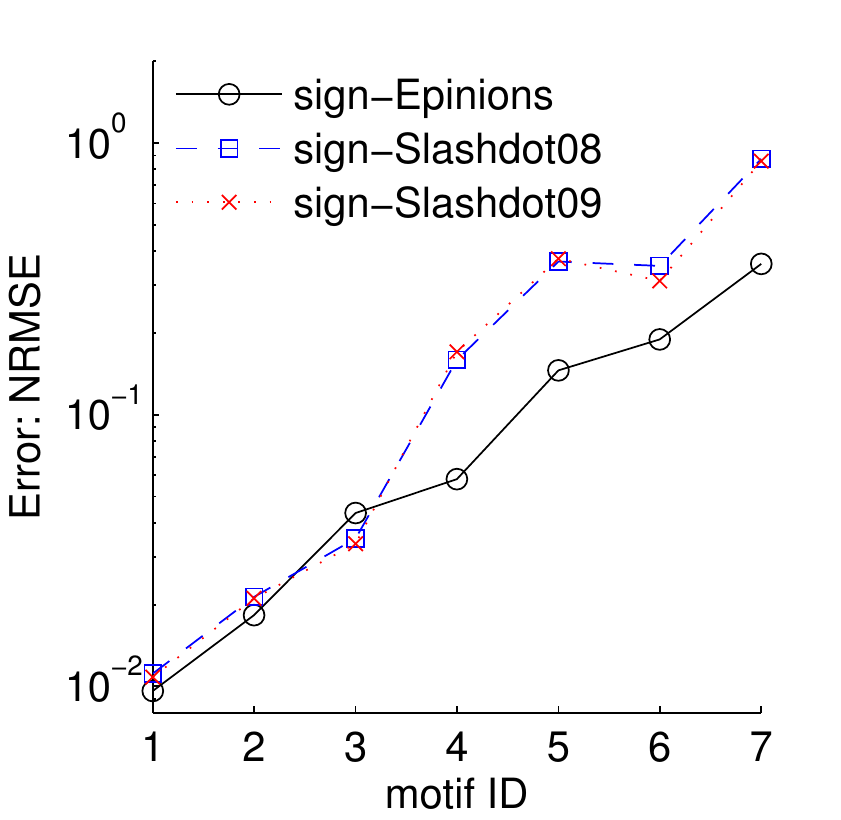}}
\subfigure[$p=0.1$]{
\includegraphics[width=0.225\textwidth]{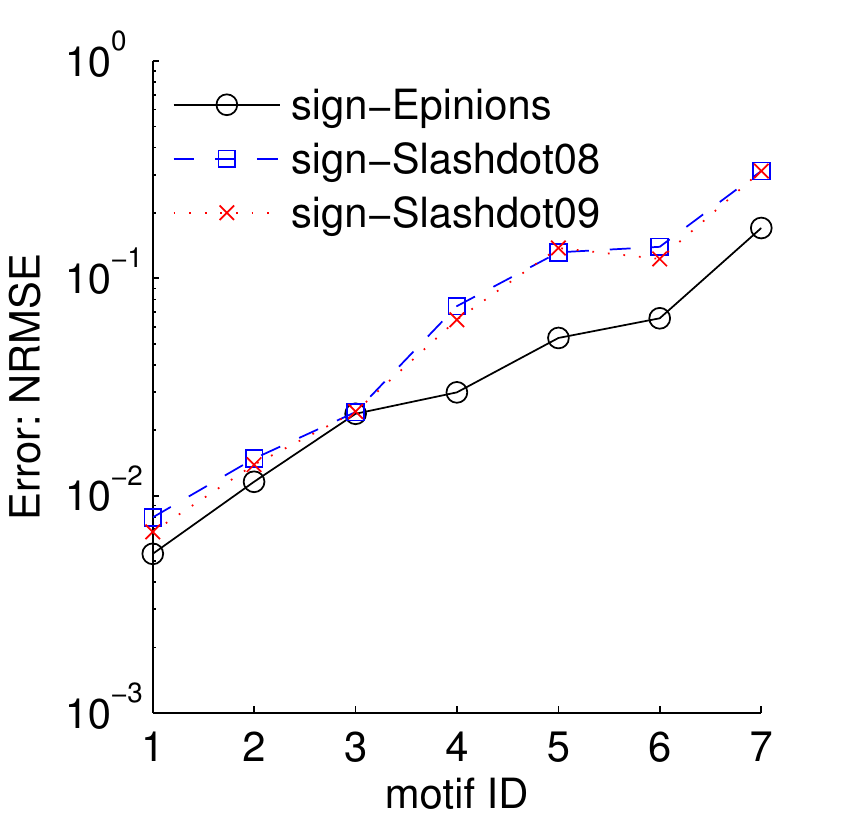}}
\caption{NRMSEs of $\omega_i^{(3)}$, the concentration estimates of 3-node signed and undirected motifs for $p=0.05$ and $p=0.1$ respectively.}
\label{fig:Results3nodeSignedCIS}
\end{figure}

Table~\ref{tab:3nodesignedgroundth} shows the real values of 3-node signed motifs' concentrations for Sign-Epinions, sign-Slashdot08, and sign-Slashdot09.
Sign-Epinions, sign-Slashdot08, and sign-Slashdot09 have $1.72\times 10^8$,  $6.72\times 10^7$, and $7.25\times 10^7$ 3-node CISes respectively.
Fig.~\ref{fig:Results3nodeSignedCIS} shows the NRMSEs of our estimates of 3-node signed and undirected motifs' concentrations for $p=0.05$ and $p=0.1$ respectively.
For all these three signed graphs, the NRMSEs are smaller than 0.9 and 0.2 when $p=0.05$ and $p=0.1$ respectively.

\subsubsection{Accuracy of inferring 4-node motifs' concentrations}
\begin{figure}[htb]
\center
%\subfigure[$p=0.05$]{
%\includegraphics[width=0.325\textwidth]{4nodeCISP05}}
\subfigure[$p=0.1$]{
\includegraphics[width=0.225\textwidth]{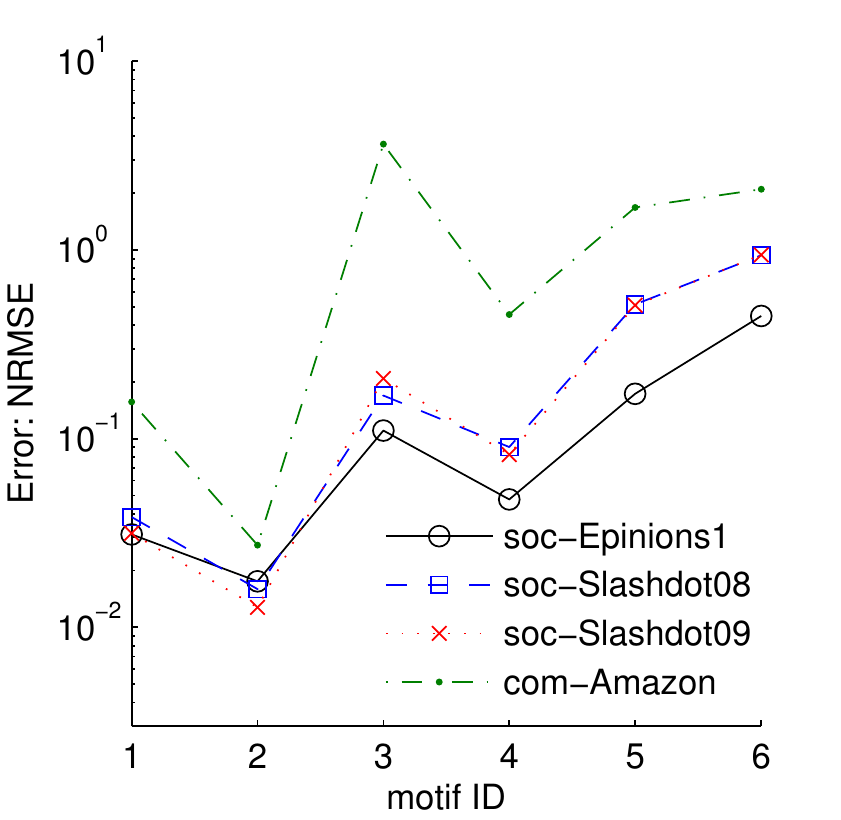}}
\subfigure[$p=0.2$]{
\includegraphics[width=0.225\textwidth]{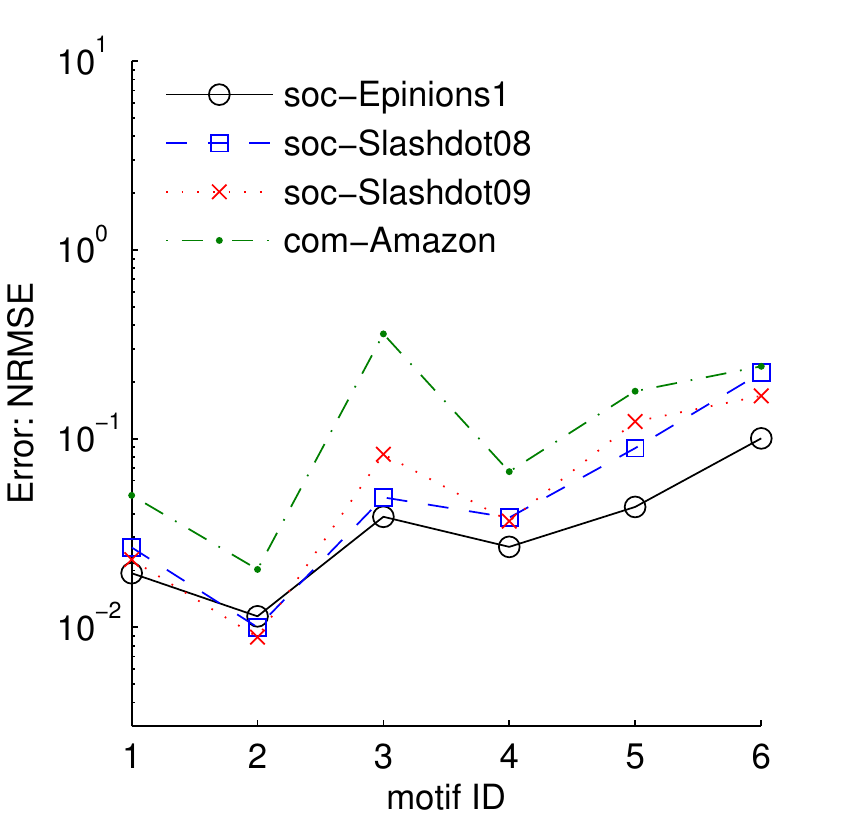}}
\caption{NRMSEs of $\hat\omega_i^{(4)}$, the concentration estimates of 4-node undirected motifs for $p=0.1$, and $p=0.2$ respectively.}
\label{fig:Results4nodeCIS}
\end{figure}

Table~\ref{tab:4nodeundergroundth} shows the real values of $\omega_i^{(4)}$, i.e., the concentrations of 4-node undirected motifs
for Soc-Epinions1, soc-Slashdot08, soc-Slashdot09, and com-Amazon.
Soc-Epinions1, soc-Slashdot08, soc-Slashdot09, and com-Amazon have $2.58\times 10^{10}$, $2.17\times 10^{10}$, $2.42\times 10^{10}$, and $1.78\times 10^8$ 4-node CISes respectively.
Fig.~\ref{fig:Results4nodeCIS} shows the NRMSEs of $\hat\omega_i^{(4)}$, the concentration estimates of 4-node undirected motifs for $p=0.05$, $p=0.1$, and $p=0.2$ respectively.
We observe that motifs with smaller $\omega_i^{(4)}$  exhibit larger NRMSEs.
Except $\omega_3^{(4)}$, the NRMSEs of the other motif concentrations' estimates are smaller than 0.2 for $p=0.2$ .

\begin{figure*}[htb]
\center
\subfigure[soc-Amazon]{
\includegraphics[width=0.325\textwidth]{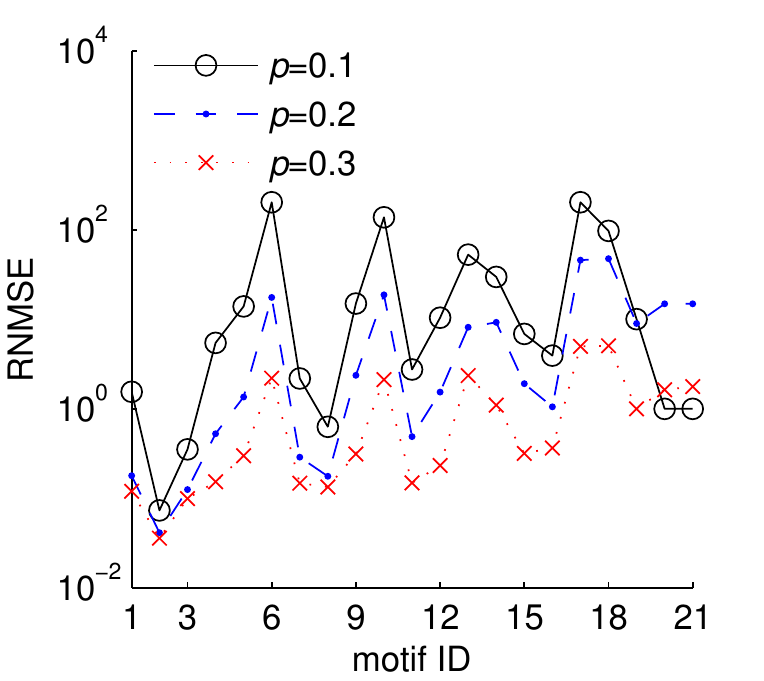}}
\subfigure[soc-DBLP]{
\includegraphics[width=0.325\textwidth]{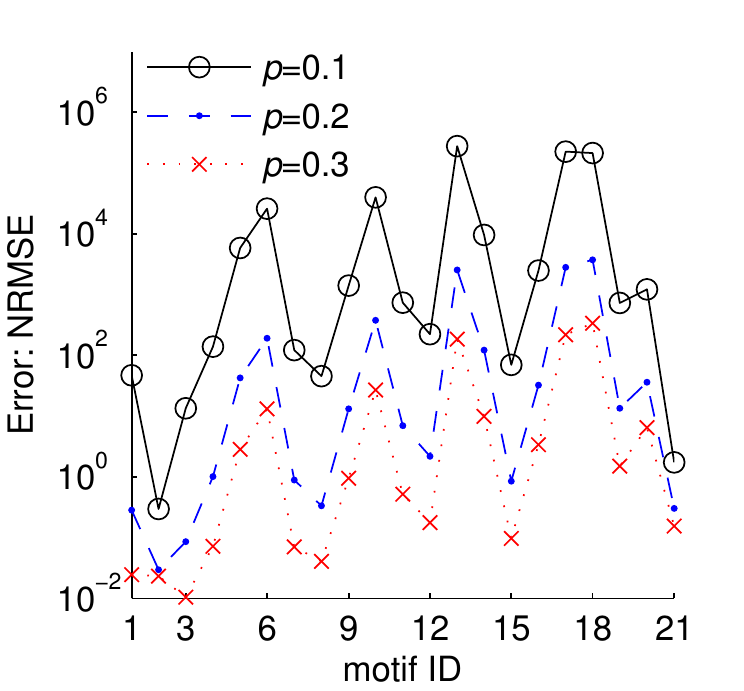}}
\subfigure[p2p-Gnutella08]{
\includegraphics[width=0.325\textwidth]{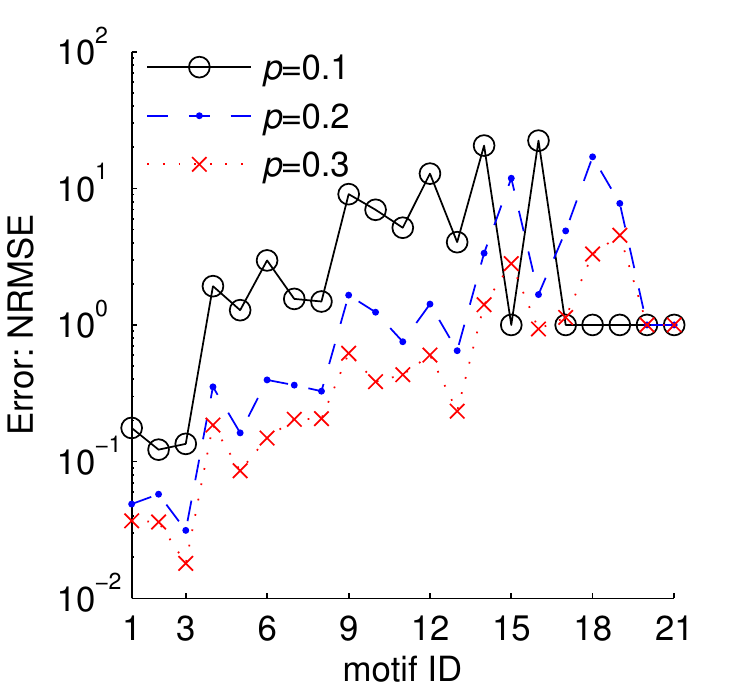}}
\subfigure[ ca-GrQc]{
\includegraphics[width=0.325\textwidth]{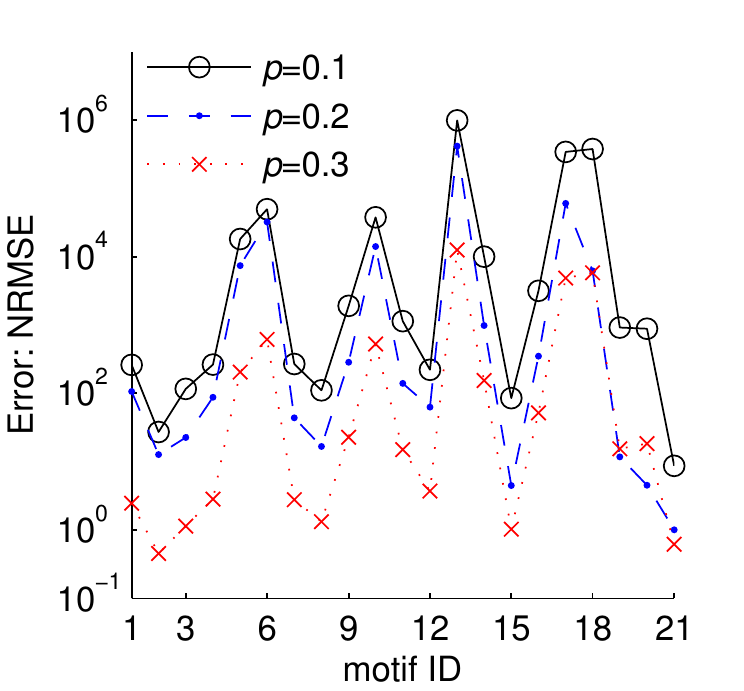}}
\subfigure[ca-CondMat]{
\includegraphics[width=0.325\textwidth]{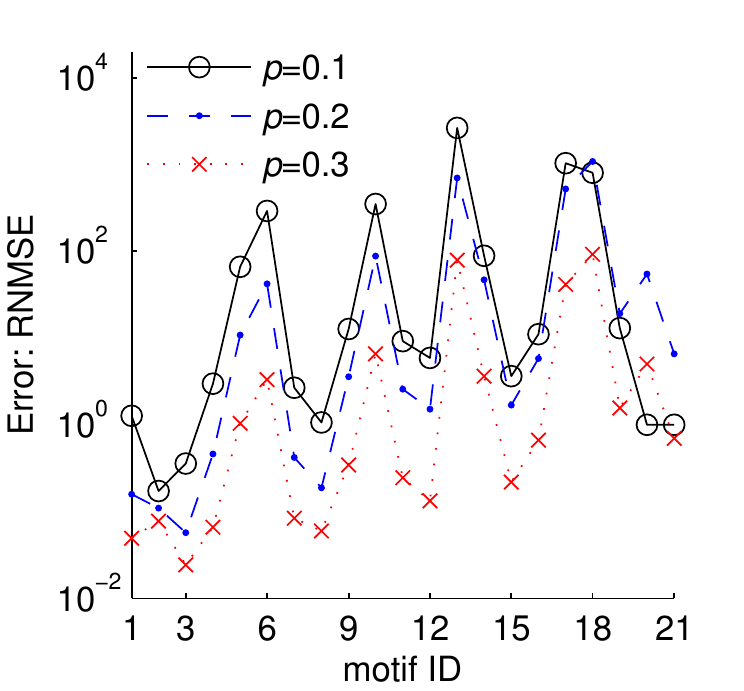}}
\subfigure[ca-HepTh]{
\includegraphics[width=0.325\textwidth]{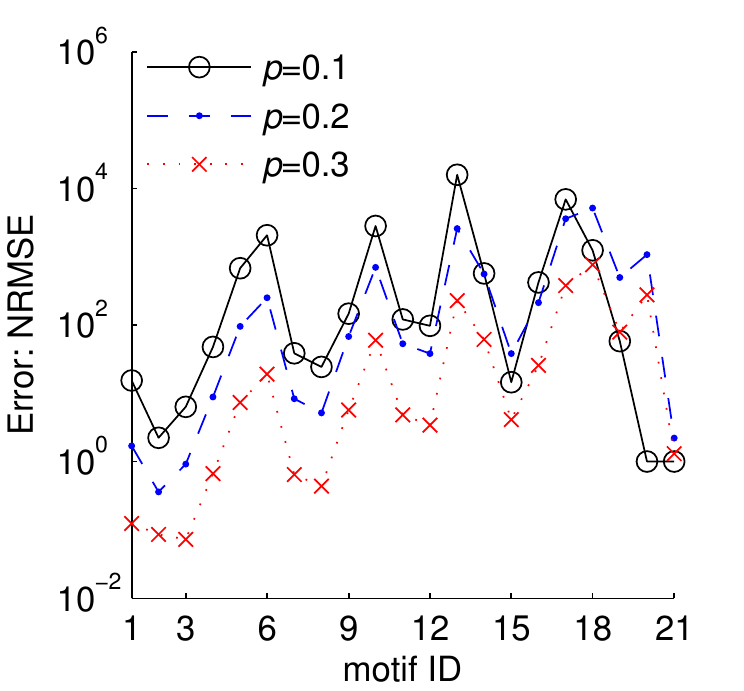}}
\caption{NRMSEs of $\hat\omega_i^{(5)}$, the concentration estimates of 5-node undirected motifs for $p=0.1$, $p=0.2$, and $p=0.3$ respectively.}
\label{fig:Results5nodeCIS}
\end{figure*}

\subsubsection{Accuracy of inferring 5-node motifs' concentrations}
Table~\ref{tab:5nodeundergroundth} shows the real values of $\omega_i^{(5)}$, i.e., the concentrations of 5-node undirected motifs
for com-Amazon, com-DBLP, p2p-Gnutella08, ca-GrQc, ca-CondMat, and ca-HepTh.
Com-Amazon, com-DBLP, p2p-Gnutella08, ca-GrQc, ca-CondMat, and ca-HepTh
contains $8.50\times 10^9$, $3.34\times 10^{10}$, $3.92\times 10^8$, $3.64\times 10^7$, $3.32\times 10^9$, and $8.73\times 10^7$ 5-node CISes respectively.
Fig.~\ref{fig:Results5nodeCIS} shows the NRMSEs of $\hat\omega_i^{(5)}$, the concentration estimates of 5-node undirected motifs for $p=0.1$, $p=0.2$, and $p=0.3$ respectively.
We observe that NRMSE decreases as $p$ increases,
and the 6-th, 10-th, 13-th, 17-th, 18-th 5-node motifs with small $\omega_i^{(5)}$  exhibit large NRMSEs.

\begin{figure*}[htb]
\center
\subfigure[(LiveJournal) 3-node directed motifs]{
\includegraphics[width=0.32\textwidth]{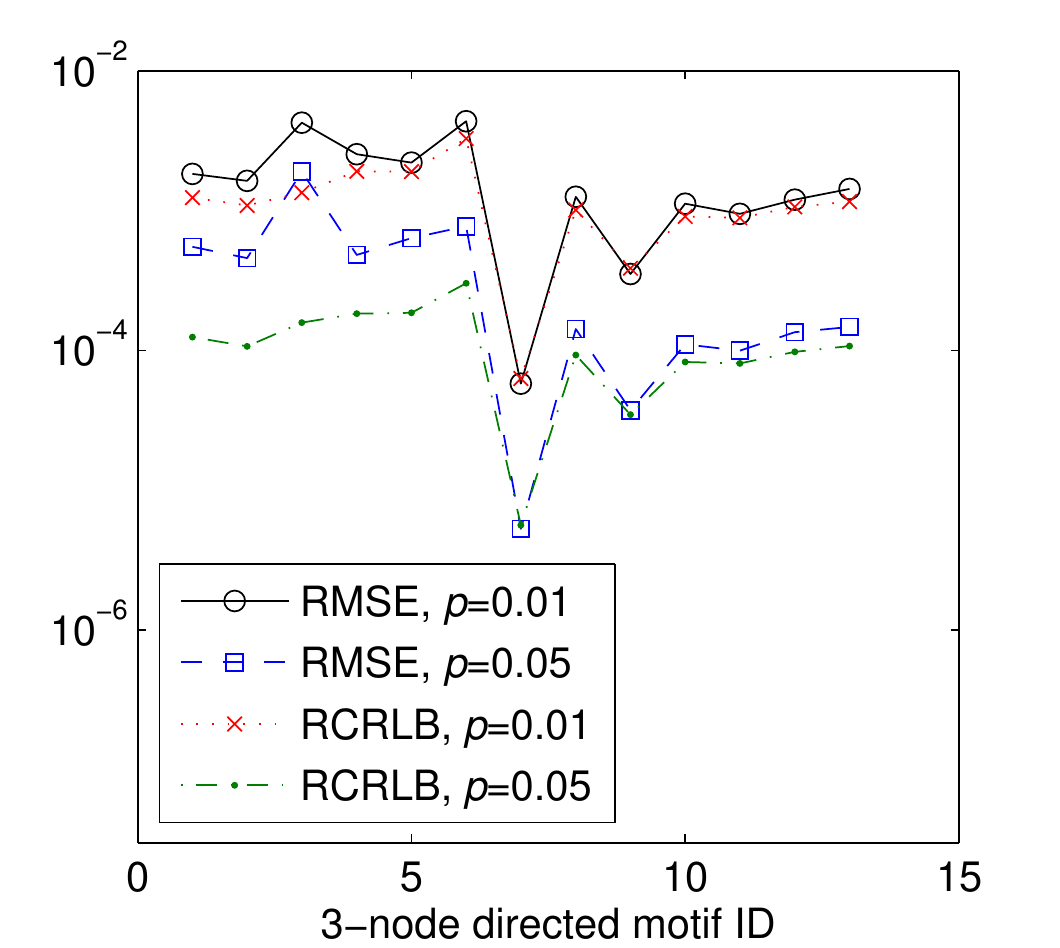}}
\subfigure[(soc-Epinions) 4-node undirected motifs]{
\includegraphics[width=0.32\textwidth]{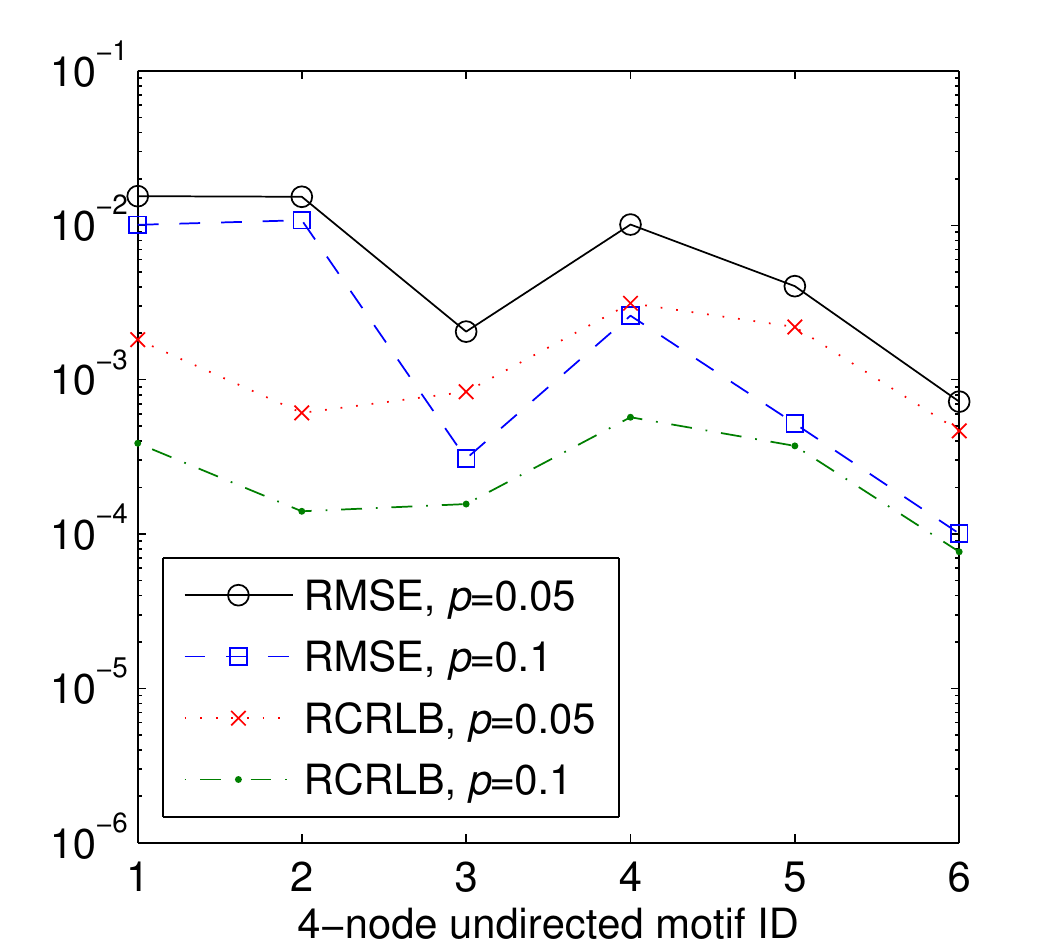}}
\subfigure[(com-DBLP) 5-node undirected motifs]{
\includegraphics[width=0.32\textwidth]{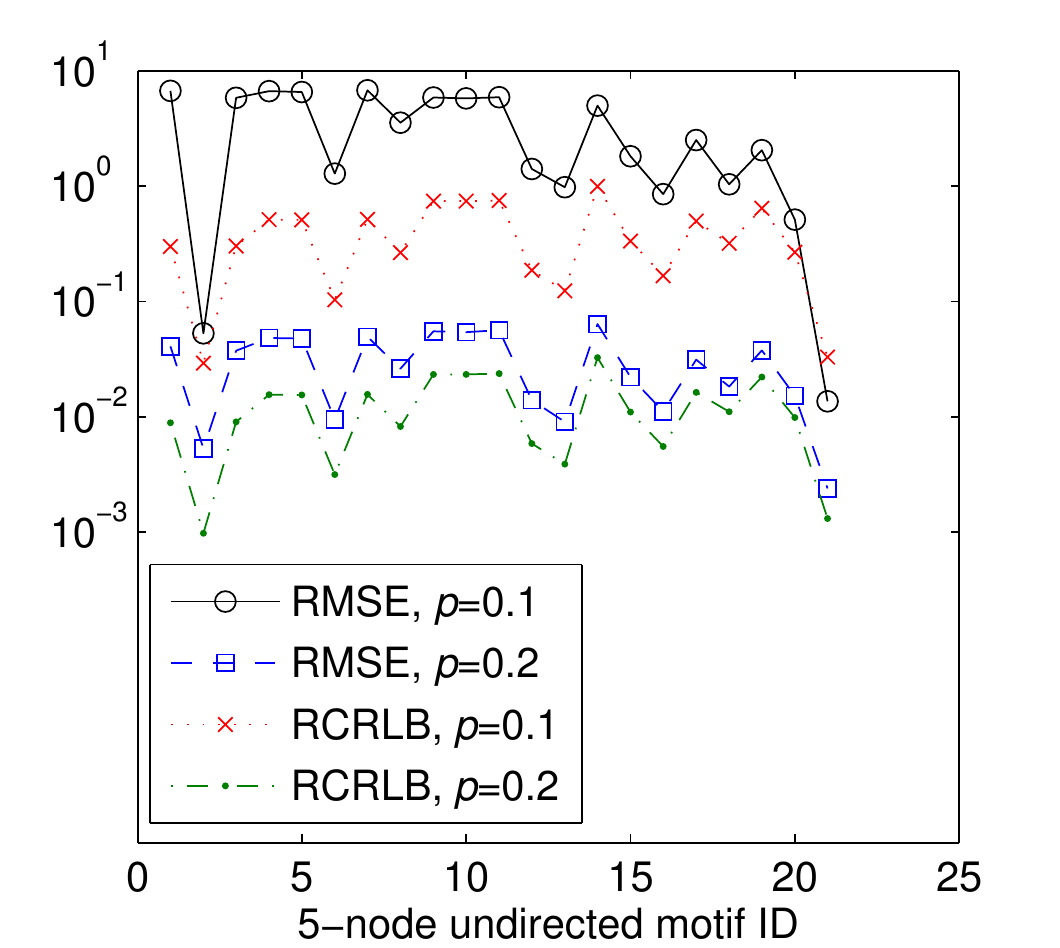}}
\caption{RCRLBs and RMSEs of concentration estimates of 3, 4, and 5-node directed motifs.}
\label{fig:CRLBnodedirected}
\end{figure*}

\subsection{Error Bounds}
Figure~\ref{fig:CRLBnodedirected} shows the root CRLBs (RCRLBs) and the root MSEs (RMSEs) of our estimates of 3-node directed motifs' concentrations, 4-, and 5-node undirected motifs' concentrations,
where graphs LiveJournal, soc-Epinions, and com-DBLP are used for studying 3-node directed motifs, 4-, and 5-node undirected motifs respectively.
We observe that the RCRLBs are smaller than the RMSEs, and fairly close to the RMSEs.
The and RCRLBs are almost indistinguishable for 3-node directed motifs, where $p=0.01$ and LiveJournal is used.
It indicates that the RCRLBs can efficiently bound the errors of our motif concentration estimations.

\section{Related Work} \label{sec:related}
There has been considerable interest to design efficient sampling methods for counting specific subgraph patterns such as
triangles~\cite{TsourakakisKDD2009,PavanyVLDB2013,JhaKDD2013,AhmedKDD2014}, cliques~\cite{ChengTODS2011,Gjoka2013}, and cycles~\cite{ManjunathESA2011}, because it is computationally intensive to compute the number of the subgraph pattern's appearances in a large graph.
Similar to the problem studied in~\cite{Kashtan2004,Wernicke2006,OmidiGenes2009,Bhuiyan2012,TKDDWang2014},
in this work we focus on characterizing 3-, 4-, and 5-nodes CISes in \emph{a single large graph},
which differs from the problem of estimating the number of subgraph patterns appearing in \emph{a large set of graphs} studied in~\cite{Hasan2009}.
OmidiGenes et al.~\cite{OmidiGenes2009} proposed a subgraph enumeration and counting method using sampling.
However this method suffers from unknown sampling bias.
To estimate subgraph class concentrations, Kashtan et al.~\cite{Kashtan2004} proposed a subgraph sampling method, but their method is computationally expensive when calculating the weight of each sampled subgraph, which is needed to correct for the bias introduced by sampling.
To address this drawback, Wernicke~\cite{Wernicke2006} proposed an algorithm, FANMOD, based on enumerating subgraph trees to detect network motifs.
Bhuiyan et al.~\cite{Bhuiyan2012} proposed a Metropolis-Hastings based sampling method GUISE to estimate 3-node, 4-node, and 5-node subgraph frequency distribution.
Wang et al.~\cite{TKDDWang2014} proposed an efficient crawling method to estimate online social networks' motif concentrations,
when the graph's topology is not available in advance and it is costly to crawl the entire topology.
In summary, previous methods focus on designing efficient sampling methods and crawling methods for estimating motif statistics when the graph is directly available or indirectly available (i.e., it is not expensive to query a node's neighbors~\cite{TKDDWang2014}).
They cannot be applied to solve the problem studied in this paper,
i.e., we assume the graph is not available but a RESampled graph is given and we aim to infer the underlying graph's motif statistics from the RESampled graph.
At last, we would like to point out our method of estimating motif statistics and its error bound computation method are inspired by methods of estimating flow size distribution for network traffic measurement and monitoring~\cite{Duffield2003,Ribeiro2006,TuneIMC08,WangTIFS2014}.

\section{Conclusions} \label{sec:conclusions}
To the best of our knowledge,
we are the first to study the problem of inferring the underlying graph's motif statistics
when the entire graph topology is not available, and only a RESampled graph is given.
We propose a model to bridge the gap between the underlying graph's and its RESampled graph's motif statistics.
Based on this probabilistic model, we develop a method Minfer to infer the underlying graph's motif statistics,
and give a Fisher information based method to bound the error of our estimates.
and experimental results on a variety of known data sets validate the accuracy of our method.

\section*{Appendix}
\begin{figure*}[htb]
\center
\subfigure[$P$ of 3-node signed subgraph classes]{
\includegraphics[width=0.43\textwidth]{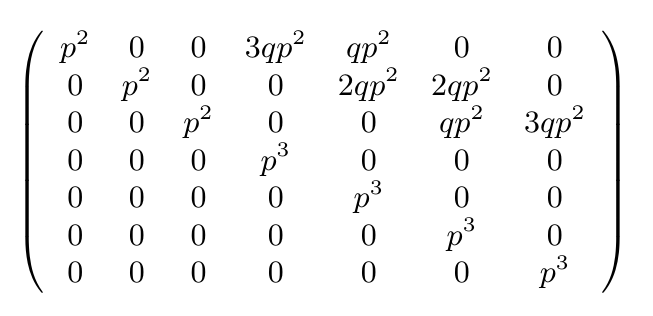}}
\subfigure[$P^{-1}$ of 3-node signed subgraph classes]{
\includegraphics[width=0.55\textwidth]{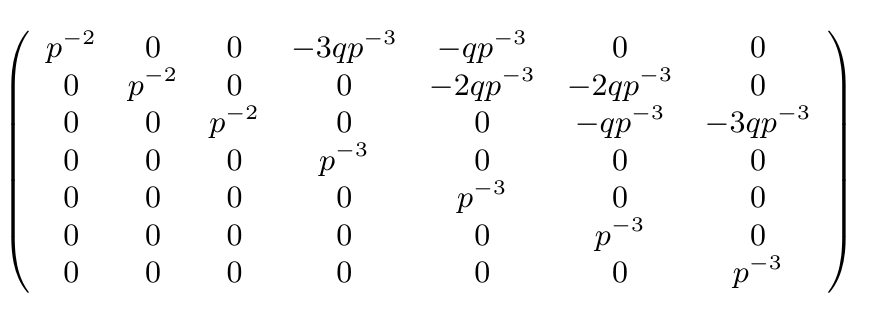}}
\subfigure[$P$ of 4-node undirected subgraph classes]{
\includegraphics[width=0.42\textwidth]{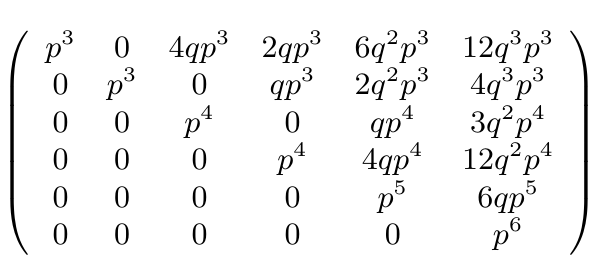}}
\subfigure[$P^{-1}$ of 4-node undirected subgraph classes]{
\includegraphics[width=0.56\textwidth]{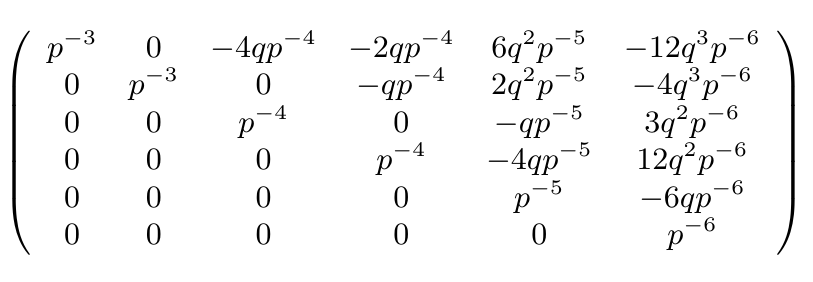}}
\subfigure[$P$ of 3-node directed subgraph classes]{
\includegraphics[width=0.72\textwidth]{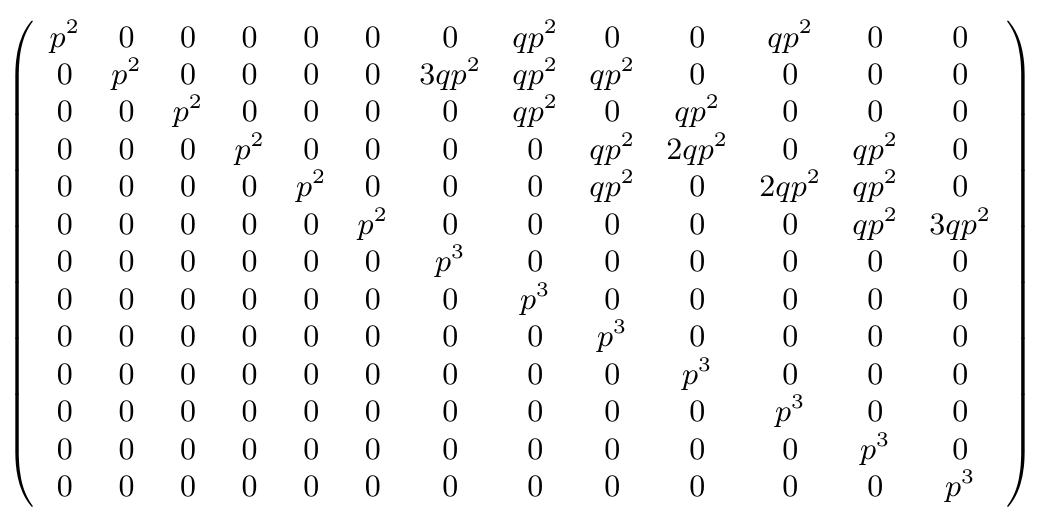}}
\subfigure[$P^{-1}$ of 3-node directed subgraph classes]{
\includegraphics[width=0.99\textwidth]{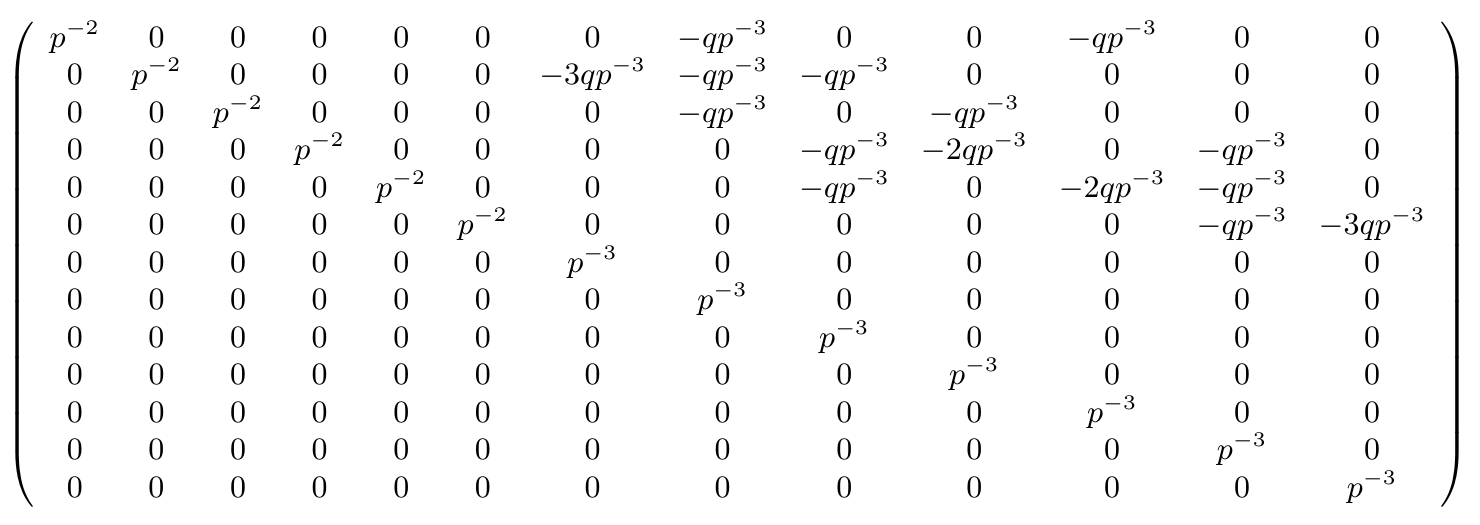}}
\caption{The matrixes $P$ and $P^{-1}$.}
\label{fig:Pij}
\end{figure*}

The matrixes $P$ and $P^{-1}$ is shown in Fig.~\ref{fig:Pij}.

\section*{Acknowledgment}
We thank the anonymous reviewers as well as Dr. Wei Fan for helpful suggestions.

\balance
\bibliographystyle{abbrv}

\end{document}